\documentclass[a4paper,12pt]{article}
\usepackage{graphicx,epsf}
\usepackage{amsbsy}

\begin{document}
\title{Nonlinear perturbations of cosmological scalar fields}
\author{David Langlois$^{1,2}$, Filippo Vernizzi$^{3,4}$\\
{\small {}}\\
{\small ${}^1${\it APC (Astroparticules et Cosmologie),}}\\
{\small {\it
UMR 7164 (CNRS, Universit\'e Paris 7, CEA, Observatoire de
Paris)}}\\
{\small {\it  B\^atiment Condorcet,  10 rue Alice Domon et L\'eonie,
 75205 Paris Cedex 13, France}}
\vspace{0.2cm}
\\
{\small ${}^2${\it GReCO, Institut d'Astrophysique de Paris, CNRS,}}\\
{\small {\it 98bis Boulevard Arago, 75014 Paris, France  }}\vspace{0.2cm}\\
{\small ${}^3${\it Helsinki Institute of Physics, P.O. Box 64,}}\\
{\small {\it FIN-00014 University of Helsinki, Finland}}\vspace{0.2cm}\\
{\small ${}^4${\it Abdus Salam International Center for Theoretical Physics}}\\
{\small {\it Strada Costiera 11, 34014 Trieste, Italy}}\\}
\date{\today}
\maketitle

\def\beq{\begin{equation}}
\def\eeq{\end{equation}}
\newcommand{\bea}{\begin{eqnarray}}
\newcommand{\eea}{\end{eqnarray}}
\def\bi{\begin{itemize}}
\def\ei{\end{itemize}}
\def\Tdot#1{{{#1}^{\hbox{.}}}}
\def\Tddot#1{{{#1}^{\hbox{..}}}}
\def\D{{\cal D}}
\def\d{{\delta}}
        \def\T{{\bf T}}
\def\u{{\partial_u}}
\def\uc{u}
\def\qc{q}
\def\e{{=}}
\def\Dc{D}
\def\varphidc{\dot{ \tilde \varphi}{}}
\def\phid{\dot \phi}
\def\chid{\dot \chi}
\def\varphid{\dot \varphi}
\def\phidc{\dot{\tilde \phi}{}}
\def\chidc{\dot{\tilde \chi}{}}
\def\L{{\cal L}}
\def\R{{\cal R}}
\def\sigmadc{\dot{\tilde \sigma}{}}
\def\sigmad{\dot \sigma}
\def\Pdc{\dot{\tilde P}{}}
\def\rhodc{\dot{\tilde \rho}{}}
\def\betadc{\dot{\tilde \beta}{}}
\def\alphadc{\dot{\tilde \alpha}{}}
\def\HH{{\cal H}}
\def\ut{\tilde u}
\def\qt{\tilde q}
\def\ht{\tilde h{}}
\def\Dt{\tilde D}
\def\hc{h}
\def\I{I}
\def\It{{(I)}}
\def\J{J}
\def\Jt{{(J)}}
\def\p{\varphi}
\def\vp{{\boldsymbol{\varphi}}}
\def\e{{\boldsymbol{e}}}

\def\Xf{X^{(1)}}
\def\Xs{X^{(2)}}
\def\af{{\delta \alpha}}
\def\as{{\delta \alpha}^{(2)}}
\def\sif{{\delta \sigma}}
\def\sis{{\delta \sigma}^{(2)}}
\def\sf{{\delta s}}
\def\ss{{\delta s}^{(2)}}
\def\phf{{\delta \varphi}}
\def\phs{{\delta \varphi}^{(2)}}
\def\phif{{\delta \phi}}
\def\chif{{\delta \chi}}
\def\sif{{\delta \sigma}}
\def\phis{{\delta \phi}^{(2)}}
\def\rhof{{\delta \rho}}
\def\rhos{{\delta \rho}^{(2)}}
\def\ps{{\delta P}^{(2)}}
\def\pf{{\delta P}}
\def\ab{\bar{\alpha}}
\def\rhob{\bar{\rho}}
\def\pb{\bar{P}}
\def\sib{\bar{\sigma}}
\def\phib{\bar{\phi}}
\def\chib{\bar{\chi}}
\def\sb{\bar{s}}
\def\thetab{\bar{\theta}}
\def\tackr{&\!\!\!}
\def\tackl{&\!\!\!}

\begin{abstract}
We present a  covariant formalism for studying  nonlinear
perturbations of scalar fields. In particular, we consider the
case of two scalar fields and introduce the notion of adiabatic
and isocurvature covectors. We obtain  differential equations
governing the evolution of these two covectors,  as well as the
evolution equation for the covector associated with the curvature
perturbation.  The form of these equations is very close to the
analogous equations obtained in the linear theory, but our
equations are fully nonlinear and exact. As an application of our
formalism, we expand these equations at second order in the
perturbations. On large scales, we obtain a closed system of
coupled scalar equations giving the evolution of the second-order
adiabatic and entropy perturbations in terms of the first-order
perturbations. These equations in general contain a nonlocal term
which, however, decays rapidly in an expanding universe.
\end{abstract}

\newpage

\section{Introduction}
In a recent series of  articles
\cite{Langlois:2005ii,Langlois:2005qp,Langlois:2006iq} we have
developed a new formalism for dealing with nonlinear cosmological
perturbations in a covariant approach. As illustrated in our
previous works, the main advantages of this new approach are its
(relative) simplicity and its broad range of validity. Indeed,
most of the earlier studies of nonlinear cosmological
perturbations rely on some approximation from the start, either by
considering quantities only up to  second order in a perturbative
approach
\cite{second_order,Acquaviva:2002ud,Maldacena:2002vr,Bartolo,Rigopoulos:2002mc,Malik:2003mv,Noh:2004bc,Vernizzi:2004nc,Tomita:2005et}
or by restricting their application to super-Hubble scales
\cite{Salopek:1990jq,Comer:1994np,Deruelle:1994iz,Sasaki:1998ug,Lyth:2003im,Rigopoulos:2003ak,Rigopoulos:2004gr,Rigopoulos:2005xx,Rigopoulos:2005,Lyth:2004gb},
or both \cite{Lyth:2005fi}. By contrast,  our approach relies on
{\em exact} equations.

Furthermore, by defining perturbations in a covariant and
geometrical way, our approach avoids many of the technical
difficulties encountered in the previous approaches, such as the
problem of the physical interpretation of second-order
perturbations in the usual coordinate-based perturbation theory.
Because we do not make any approximation,  it is also rather
straightforward to make the connection with the other approaches.

In our previous papers we have focused our attention on fluids,
either perfect or dissipative. The purpose of the present work is
to extend our formalism to scalar fields, as they
play an essential role in early universe models.

An elegant covariant approach to treat {\em perfect fluid}
inhomogeneities about a Friedmann-Lema\^{\i}tre-Robertson-Walker
(FLRW) spacetime was  introduced in \cite{Ellis:1989jt}, based on
earlier work by Hawking  \cite{Hawking:1966qi}. This approach was
later  extended to scalar field dominated universes in
\cite{Bruni:1991kb}, although   essentially by describing the
scalar field as an effective fluid.

Here we consider the situation where several scalar fields are
present, concentrating on the case of {\em two} scalar fields. In
the latter case, it has been shown {\em in the linear limit} that
one can define so-called adiabatic and entropy components for the
perturbations by a suitable rotation of the original scalar field
perturbations \cite{Gordon:2000hv,bartjan}. These two components
obey a coupled system of second-order (in time) differential
equations, which decouple when the background evolution in field
space corresponds to a straight trajectory. This decomposition can
also be useful for analyzing the generation of adiabatic and entropy
perturbations during inflation, which in general can be
correlated, as  first pointed out in \cite{Langlois:1999dw}.

A {\em nonlinear} formalism for scalar fields with a similar
decomposition into adiabatic and entropy components has been
developed in \cite{Rigopoulos:2005xx}, making use of the long
wavelength approximation, which consists
in dropping higher-order terms in the spatial gradients,
assuming that these are negligible on {\em large scales}
\cite{Salopek:1990jq,Comer:1994np,Deruelle:1994iz}. In  the
present work, we show that it is possible to generalize this
decomposition into adiabatic and isocurvature components to the
exact and fully nonlinear situation, without resorting to any
approximation. In order to do so, we
introduce adiabatic and isocurvature {\em covectors}, which
generalize the linear definitions. We then derive a coupled system
of second-order (with respect to time) equations for these two
components  which are exact and covariant, and mimic the equations
of the linear theory.

The covariant equations that we obtain can be expanded at first
and second order in the perturbations by working in some
coordinate system. At first order in the perturbations, it is
straightforward to recover the results of \cite{Gordon:2000hv},
i.e., two coupled evolution equations for the adiabatic and
entropy perturbations and an evolution equation for the curvature
perturbation on comoving or uniform density hypersurfaces. At
second order in the perturbations, we extend the results of
\cite{Vernizzi:2004nc,Malik:2005cy} and we derive, on {\em large
scales}, an evolution equation for the (second-order) entropy
perturbation, which sources the evolution of the (second-order)
curvature perturbation.

Furthermore, we will show that the second-order two-field system
has qualitatively new features with respect to the linear case or
the second-order single-field case. In particular, the scalar
equations that we derive contain generically a nonlocal term due
to the impossibility of writing the total momentum of the two
fields as a total gradient. However, we will show that this
nonlocal term can be neglected, on large scales, due to the
expansion of the universe.

This work is organized as follows: in the next section we present
the basic equations to treat a system of  several scalar fields in
the covariant formalism. In Sec.~\ref{sec:single} we discuss the
perturbations of a single-field dominated universe, while in
Sec.~\ref{sec:two} we extend this treatment to two fields. In
Sec.~\ref{sec:approximate} we give a set of covariant equations
that describe the two-field system in two limits: the {\em linear}
regime and the large-scale regime. From our covariant equations we
derive, in the coordinate approach, the evolution equations for
the linear perturbations in Sec.~\ref{sec:linear} and for the
second-order perturbations on large scales in Sec.
\ref{sec:secondorder}. Finally, in Sec.~\ref{sec:conclusion} we
draw our conclusions.

\section{Covariant formalism for several scalar fields}
\label{sec:covariant}

Let us consider an arbitrary unit timelike vector $u^a= d
x^a/d \tau$ ($u_a u^a =-1$), which defines a congruence of
cosmological observers. The spatial projection tensor orthogonal
to the four-velocity $u^a$ is defined by
\beq
h_{ab}\equiv g_{ab}+u_a u_b, \quad \quad (h^{a}_{\ b} h^b_{\
c}=h^a_{\ c}, \quad h_a^{\ b}u_b=0).
\eeq
To describe the time evolution, {\em our} covariant definition of
the time derivative  throughout this work will be the {\em Lie derivative}
with respect to $u^a$, which is defined for a generic covector
$X_a$ by (see e.g. \cite{wald})
\beq
\dot X_a\equiv\L_u X_a \equiv u^b \nabla_b X_a + X_b \nabla_a u^b,
\label{Lie}
\eeq
and will be denoted by {\em a dot}. We emphasize that this
notation  differs from the convention used in the literature based
on the covariant formalism (see e.g. \cite{Ellis:1989jt}), as well
as in our previous papers
\cite{Langlois:2005ii,Langlois:2005qp,Langlois:2006iq}, where a
dot denoted simply the covariant derivation along $u^a$. The two
definitions coincide only for scalar  quantities,
\beq
\dot f = u^b \nabla_b f.
\eeq

To describe perturbations in the covariant approach, we introduce
the projected covariant derivative orthogonal to the four-velocity
$\uc^a$, $\Dc_a$, in the spirit of \cite{Ellis:1989jt}. For a
generic tensor, the definition is
\beq
D_a T_{b\dots}^{\ c\dots}\equiv h_a^{\ d}h_b^{\ e}\dots h^{\
c}_f\dots\nabla_d T_{e\dots}^{\ f\dots}.
\eeq
In particular, for a scalar quantity $f$,
the definition reduces to
\beq
\Dc_a f\equiv h_a^{\ b} \nabla_b f = \partial_a f + u_a \dot f \, .
\label{def_Daf}
\eeq
We also introduce the familiar decomposition
\beq
\nabla_b u_a=\sigma_{ab}+\omega_{ab}+{1\over 3}\Theta h_{ab}-a_a
u_b, \label{decomposition}
\eeq
with the (symmetric) shear tensor $\sigma_{ab}$ and the
(antisymmetric) vorticity  tensor $\omega_{ab}$; the volume
expansion, $\Theta$, is defined by
\beq
\Theta \equiv \nabla_a u^a,
\eeq
while
\beq
a^a \equiv u^b\nabla_b u^a
\eeq
is the acceleration vector.

Let us now  consider $N$ scalar fields
minimally coupled to gravity with Lagrangian density
\beq
{\cal L} = - \frac{1}{2}
\partial_a \p_\I \partial^a \p^\I -  V(\p_K) ,
\label{lagrangian}
\eeq
where $V$ is the potential. We assume, for simplicity,
canonical kinetic terms. Here and in the following, summation
over the field indices ($I,J,\ldots$)  will be implicit. The
energy-momentum tensor derived from this Lagrangian reads
\beq
T_{ab}= \partial_a \p_\I \partial_b \p^I -\frac{1}{2} g_{ab}
\left( \partial_c \p_\I
\partial^c \p^I + 2 V \right). \label{EMT}
\eeq
Given an arbitrary unit timelike vector field $u^a$, it is always
possible to decompose the total energy momentum tensor as
\beq
T_{ab} =(\rho+P) u_a u_b  + q_au_b+ u_aq_b+ g_{ab} P + \pi_{ab}, \label{EMT2}
\eeq
where $\rho$, $P$, $q_a$ and $\pi_{ab}$ are respectively the energy density,
pressure,  momentum  and anisotropic stress tensor measured in the
frame defined by $u^a$.
Starting from the energy-momentum tensor (\ref{EMT}) one finds
\bea
\rho \tackl \equiv \tackr 
T_{ab}u^au^b= \frac{1}{2} \left(   \dot \p_I\dot\p^I + \Dc_a
\p^I \Dc^a \p_\I \right) +V , \label{rhotot}\\
P \tackl \equiv \tackr \frac{1}{3}h^{ac}T_{ab}h^b_{\, c}= \frac{1}{2}  \left(
\dot \p_I\dot\p^I - \frac{1}{3} \Dc_a \p_\I
\Dc^a \p^I \right) -V, \label{Ptot}\\
q_a \tackl \equiv \tackr 
-u^bT_{bc}h^c_{\, a}=  - \dot \p_\I \Dc_a \p^\I, \label{q}\\
\pi_{ab} \tackl \equiv \tackr   h_a^{\, c}T_{cd} h^d_{\, b}-P h_{ab}= \Dc_a
\p_\I \Dc_b \p^I -\frac{1}{3} \hc_{ab} \Dc_c \p_\I \Dc^c \p^\I .
\label{as}
\eea

The evolution equations for the scalar fields are obtained from the variation
of the action with respect to the scalar fields. One thus gets $N$ Klein-Gordon
equations, given by
\beq
\nabla_a\nabla^a\p_\I=\frac{\partial V}{\partial \p_\I}.
\eeq
Like for the energy-momentum tensor, it is useful to consider a decomposition
into (covariant) time-like and space-like gradients defined with respect
to $u^a$. One then finds that the above Klein-Gordon equations can
be re-expressed in the form
\beq
\ddot \p_\I + \Theta \dot \p_\I +V_{,\varphi_\I}-\Dc_a \Dc^a \p_\I - a^a
\Dc_a \p_\I =0, \label{evolp}
\eeq
where we have defined
\beq
V_{,\varphi_\I} \equiv \frac{\partial V}{\partial \p_\I}.
\eeq
In the following sections, we will consider in turn the case of a single scalar field, which has already been
studied briefly in our covariant approach
\cite{Langlois:2005ii}, and the case of two scalar fields.

\section{Single field}
\label{sec:single}

\subsection{Equation of motion}

For a single scalar field, which will be denoted by $\phi$, the
Klein-Gordon equation (\ref{evolp}) reads
\beq
\ddot \phi + \Theta \dot \phi +\frac{d V}{d\phi}-\Dc_a \Dc^a \phi
- a^a \Dc_a \phi =0. \label{evol_single}
\eeq
Although this equation is fully nonlinear, it closely mimics the
{\it homogeneous} Klein-Gordon equation in a FLRW spacetime when
one identifies $\Theta/3$ with the  local Hubble parameter. However, the
last two terms depend explicitly on the spatial gradients
but it is possible, in the case of a single-field, to get rid of
them by a particular choice of the unit vector $u^a$. Indeed, if
the gradient $\nabla_a\phi$ is time-like,  which is the case if
the spacetime is sufficiently close to FLRW, then one can always
choose for $u^a$ the unit vector field orthogonal to the $\phi$
hypersurfaces,
% change FV 08-01-2006
\beq
u^a_{\rm com}=\pm
\frac{\nabla^a \phi}{\sqrt{-\nabla_b\phi\nabla^b\phi}},  \label{u_comoving}
\eeq
% end of change
where the sign of the right hand side depends whether $\nabla_a\phi$ is
future or past oriented. In this {\em comoving} frame, the spatial
gradient of $\phi$, i.e.~$\Dc_a\phi$, automatically vanishes.
Therefore, for this particular choice of $u^a$, the last two terms
on the left hand side of (\ref{evol_single}) disappear and the
Klein-Gordon equation looks formally similar to the homogeneous
version although it remains fully inhomogeneous and nonlinear.

One can go one step further by deriving an exact and covariant
equation that mimics the equation of motion governing the linear
perturbations of a scalar field in a perturbed FLRW spacetime. The
trick  is to  consider  the gradient  of the
scalar field, i.e.,  the  covector
\beq
\phi_a \equiv \nabla_a \phi.  \label{phi_a_def}
\eeq
This can be decomposed into a spatial gradient and a
longitudinal component,
\beq
\phi_a = \Dc_a\phi-\dot \phi\  u_a.
\eeq
Note that with the choice (\ref{u_comoving}) the
spatial gradient disappears in the above expression and this
equation can be used to rewrite the comoving four-velocity
as $u^a_{\rm com} = - \nabla^a \phi/\dot \phi$.

We now construct a second-order (in time) evolution equation for
$\phi_a$. We recall  that a dot stands for  the Lie derivative
along $u^a$, as defined in Eq.~(\ref{Lie}). We can derive the
evolution equation for $\phi_a$ by taking the
 spacetime gradient of Eq.~(\ref{evol_single}) and noting that, for
any scalar $\phi$,  the Lie derivative with respect to $u^a$ and
the spacetime gradient (but not the spatial gradient) commute,
i.e.,
\beq
\nabla_a \dot \phi = \Tdot{(\nabla_a \phi)}.
 \label{prop_Lie}
\eeq
Equation (\ref{evol_single}) then yields
\beq
\label{eq_phi_a} \ddot \phi_a + \Theta \dot \phi_a + V_{,\phi
\phi} \phi_a= -  \dot \phi \nabla_a
\Theta+\nabla_a(\Dc^b\Dc_b\phi)+\nabla_a(a^b\Dc_b\phi).
\eeq
This equation mimics the analogous perturbation equation at linear
order but also  incorporates the fully nonlinear dynamics of the
scalar field perturbation.

\subsection{Integrated expansion perturbation on co\-mo\-ving sli\-ces}

In \cite{Langlois:2005ii,Langlois:2005qp} it was shown that one
can define a covariant generalization of  the comoving
curvature perturbation by using appropriate combinations of
spatially projected gradients. For a scalar field, a natural
choice is the covariant integrated expansion perturbation on
comoving hypersurfaces
$\R_a$, defined as
\beq
\R_a \equiv -D_a \alpha + \frac{\dot \alpha}{\dot \phi} D_a \phi,
\label{R_def}
\eeq
where $\alpha$ is the integrated volume expansion along $u^a$,
\beq
\alpha \equiv {1\over 3}\int d\tau \, \Theta \quad \quad (\Theta
= 3 \dot \alpha  ) \label{alpha_def}.
\eeq
 Since $\Theta/3$ corresponds to the local Hubble parameter, one sees that the quantity $\alpha$ can
be interpreted as the number of e-folds measured along the world-line of a cosmological observer with
four-velocity $u^a$.

When the four-velocity $u^a$ is chosen to be comoving with the scalar field, as defined in Eq.~(\ref{u_comoving}), then
the last term in the definition (\ref{R_def}) drops out. For this particular case,
the evolution equation for $\R_a$ has already been given in
Ref.~\cite{Langlois:2005ii,Langlois:2005qp}.

Note a useful  property of  $\R_a$: one can
replace in its definition (\ref{R_def})
the spatial gradients $D_a$ by partial or covariant
derivatives,
\beq
\R_a = -\nabla_a \alpha + \frac{\dot \alpha}{\dot \phi} \phi_a.
\eeq
The same property applies to  the nonlinear
generalization of the comoving Sasaki-Mukhanov variable for a
scalar field, which can be defined as
\beq
Q_a \equiv D_a \phi - \frac{\dot \phi}{\dot \alpha} D_a \alpha
=\frac{\dot \phi}{\dot \alpha}\R_a. \label{Q_single}
\eeq

\subsection{Integrated expansion perturbation on uniform energy density slices}

It is also possible to generalize the curvature perturbation on
uniform energy density hypersurfaces, as shown in detail in our
previous works \cite{Langlois:2005ii,Langlois:2005qp}. The key
role is then played by the covector $\zeta_a$ defined as
\beq
\zeta_a \equiv D_a \alpha - \frac{\dot \alpha}{\dot \rho} D_a
\rho. \label{zeta}
\eeq
For a perfect fluid, the quantity $\zeta_a$ satisfies a remarkably
simple first-order evolution equation
\cite{Langlois:2005ii,Langlois:2005qp},
\beq
\dot \zeta_a = \frac{\Theta^2}{3 \dot \rho} \Gamma_a,
\label{zeta_evolution_single}
\eeq
(the dot stands, as before, for a  Lie derivative with respect to $u^a$) where
\beq
\Gamma_a \equiv D_a P -\frac{\dot P}{\dot \rho} D_a \rho
\label{Gamma_def}
\eeq
is the nonlinear nonadiabatic pressure perturbation. For a
barotropic fluid, $\Gamma_a=0$ and $\zeta_a$ is conserved on {\em
all scales}.  The relation (\ref{zeta_evolution_single}) for
$\zeta_a$ can be seen as a generalization of the familiar
conservation law for $\zeta$, the linear curvature perturbation on
uniform energy hypersurfaces. It can also be extended to a
non-perfect fluid \cite{Langlois:2006iq}.

Note however that the relation between our $\zeta_a$ and the traditional
$\zeta$ is somewhat subtle on small scales. As we showed explicitly in
\cite{Langlois:2005qp}, if we work in some coordinate system the spatial
components of $\zeta_a$ correspond essentially to the spatial
gradient of $\zeta$ on super-Hubble scales, but on smaller scales there is
a non-negligible  extra term. Consequently, on small scales Eq.~(30)
does not imply, for adiabatic perturbations, the conservation of the
traditional $\zeta$.

For a scalar field, the comoving and uniform density
integrated expansion perturbations $\zeta_a$ and
$\R_a$ satisfy
\beq
\zeta_a + \R_a = - \frac{\dot \alpha}{\dot \rho} \left( D_a \rho -
\frac{\dot \rho}{\dot \phi} D_a \phi \right),
\eeq
which simply follows from their respective definitions. The right
hand side can be interpreted as the ``shift'' between
hypersurfaces of constant $\rho$ and hypersurfaces of constant
$\phi$ and the term inside the parenthesis represents the
nonlinear generalization of the so-called {\em comoving energy
density} of a single scalar field.

By choosing $u^a=u^a_{\rm com}$  as defined in Eq.~(\ref{u_comoving}), 
the energy-momentum
tensor of a single scalar field can be written in
the perfect fluid form, i.e., with vanishing $q_a$ and $\pi_{ab}$ 
and the energy density $\rho$ and
pressure $P$ given by
\beq
\rho= \frac{1}{2} \dot \phi^2 + V, \qquad P= \frac12 \dot \phi^2 -
V, \label{rho_P_single}
\eeq
as can be checked by specializing Eqs.~(\ref{rhotot}--\ref{as}) to a
single field and setting $D_a \phi=0$.

The scalar field cannot, strictly speaking, be considered as a
barotropic fluid. Indeed, Eq.~(\ref{rho_P_single}) implies
\beq
P=\rho-2 V \label{Ptorhosingle},
\eeq
which, after substitution in  the definition (\ref{Gamma_def}) of
the nonadiabatic pressure covector $\Gamma_a$, gives
\beq
\Gamma_a = \left(1 - \frac{\dot P}{ \dot \rho} \right) D_a \rho,
\label{gamma_single}
\eeq
where we have used $D_a V=0$. Substituting  the  time
derivative of Eq.~(\ref{Ptorhosingle}), one gets
\beq
\Gamma_a =2  \frac{\dot \phi}{\dot \rho} V_{,\phi}D_a \rho,
\eeq
and the evolution equation of $\zeta_a$ for a single
scalar field is then given by
\beq
\dot \zeta_a = \frac{2}{3}  \frac{V_{,\phi}}{ \dot \phi^3} D_a \rho,
\eeq
where we have used $\dot\rho=-\Theta\dot\phi^2$ to obtain this
expression. As we will show in Sec.~\ref{sec:ls},
 the right hand side of this equation can be neglected
on large scales, so that $\zeta_a$ is conserved in this limit.

%%%%%%%%%%%%%%%%%%%%%%%%%%%%%%%%%%%%%%%%%%%%%%%%%%%%%%%%%%%%%%%%%%%%%%%%%%%%%
%%%%%%%%%%%%%%%%%%%%%%%%%%%%%%%%%%%%%%%%%%%%%%%%%%%%%%%%%%%%%%%%%%%%%%%%%%%%%
\section{Two scalar fields}
\label{sec:two}
%%%%%%%%%%%%%%%%%%%%%%%%%%%%%%%%%%%%%%%%%%%%%%%%%%%%%%%%%%%%%%%%%%%%%%%%%%%%%
%%%%%%%%%%%%%%%%%%%%%%%%%%%%%%%%%%%%%%%%%%%%%%%%%%%%%%%%%%%%%%%%%%%%%%%%%%%%%

We now consider several scalar fields and for simplicity we
restrict our analysis to the case of two scalar fields, which we
will denote by $\phi$ and $\chi$.

\subsection{Adiabatic and entropy covectors}

In the two-field case, it is possible to introduce a particular
basis in the field space in which various field dependent
quantities are decomposed into so-called adiabatic and entropy
components. In the linear theory, this decomposition
 was first introduced in
\cite{Gordon:2000hv} for two fields. For the multi-field case, it
is discussed in \cite{bartjan} in the linear theory and in
\cite{Rigopoulos:2005xx} in the nonlinear context.

In our case, the corresponding basis consists, in the
two-dimensional field space, of   a unit vector $\e_\sigma^I$
defined in the direction of the velocity of the two fields, and thus
{\em tangent} to the trajectory in field space, and of a  unit
vector $\e_s^I$ defined along  the direction {\em orthogonal} to
it, namely
\beq
\e_\sigma^\I \equiv  \frac{1}{\sqrt{\dot\phi^2+\dot\chi^2}}
\left(\dot \phi, \dot \chi \right),
\qquad \e_s^\I \equiv
 \frac{1}{\sqrt{\dot\phi^2+\dot\chi^2}}
\left(- \dot \chi, \dot \phi \right)
 \label{e1} \label{e2}.
\eeq
An immediate consequence of the above definitions is the identity
\beq
\delta_\J^\I = \e_\sigma^\I \e_\sigma{}_\J
+ \e_s^\I \e_s{}_\J \, , \label{projector}
\eeq
which will be  very useful in the derivation of many relations.

To make the notation shorter it is  convenient
to introduce the  angle $\theta$ defined by
\beq
\cos \theta \equiv \frac{\phid}{\sqrt{\dot\phi^2+\dot\chi^2}}\, ,
\quad \quad \sin \theta \equiv
\frac{\chid}{\sqrt{\dot\phi^2+\dot\chi^2}}\, ,
\eeq
so that
\beq
\e_\sigma^\I =
\left(\cos \theta,\sin \theta  \right),
\qquad \e_s^\I =
\left(-\sin \theta , \cos \theta \right).
\eeq
This angle, in contrast with the linear theory case where it is a
background quantity that depends only on time,  is here an
inhomogeneous  quantity which depends on  time and space. By
taking the time derivative of the basis vectors $\e_\sigma^\I$ and
$\e_s^\I$, we get
\beq
\dot \e_\sigma^\I =  \dot \theta \e_s^\I  , \qquad
\dot \e_s^\I = - \dot \theta
\e_\sigma^\I. \label{angle_1}
\eeq
It is also convenient to introduce the {\em formal} notation
\beq
\sigmad\equiv\sqrt{\dot\phi^2+\dot\chi^2}.
\eeq
Note however that this notation can be misleading as, in
general, in the nonlinear context, $\sigmad$ {\em is not} the derivative along $u^a$ of a
scalar field $\sigma$.

Making use of the basis (\ref{e1}), one can then introduce  two
linear combinations of the scalar field gradients and thus define
two covectors, respectively denoted by $\sigma_a$ and $s_a$, as
\bea
\sigma_a \tackl \equiv \tackr \e_\sigma^\I \nabla_a \p_\I=\cos\theta\, \nabla_a\phi+\sin\theta\, \nabla_a\chi
\label{tan_ort1}, \\
 s_a \tackl \equiv \tackr \e_s^\I \nabla_a \p_\I  =-\sin\theta\, \nabla_a\phi+\cos\theta\, \nabla_a\chi
 \label{tan_ort2}.
\eea
We will call these two covectors the {\em adiabatic} and {\em
entropy}  covectors, respectively, by analogy with the similar
definitions in the linear context \cite{Gordon:2000hv}. Whereas
the entropy covector $s_a$ is orthogonal to the four-velocity
$u^a$, i.e., $u^a s_a = 0$, this is not the case for $\sigma_a$
which contains a ``longitudinal'' component:
$u^a\sigma_a=\dot\sigma$. It is also useful to introduce the
spatially projected version of (\ref{tan_ort1}-\ref{tan_ort2}),
\def\sp{\sigma^{_\perp}}
\beq
\sp_a \equiv \e_\sigma^\I D_a \p_\I = \sigma_a+
\dot\sigma u_a \, , \qquad
s^\perp_a \equiv
\e_s^\I D_a \p_\I = s_a\, .
\label{perp}
\eeq

Let us now consider the ``adiabatic'' combination of the
Klein-Gordon equations, i.e., the contraction of (\ref{evolp}) by
$\e_\sigma^\I$. By noting that
\beq
\e_\sigma^\I \dot \p_\I=\sigmad, \qquad \e_\sigma^\I \ddot
\p_\I=\ddot\sigma,
\eeq
and by defining
\beq
V_{, \sigma} \equiv \e_\sigma^\I V_{,\p_\I},
\eeq
one obtains
\beq
\ddot \sigma + \Theta \dot \sigma +V_{,\sigma} -\e_\sigma^\I \Dc_a
\Dc^a \p_\I -a^a\sp_a=0.
\eeq
The fourth term can be rewritten as
\beq
\e_\sigma^\I \Dc_a \Dc^a \p_\I=D_a \left(\e_\sigma^\I  \Dc^a
\p_\I\right)- (D_a\e_\sigma^\I ) \Dc^a \p_\I= D^a\sp_a-(\e_{s\I}
D_a\e_\sigma^\I) s^a,
\eeq
where we have used the definition of $\sigma_a^\perp$, Eq.~(\ref{perp}),
 and the identity (\ref{projector}).
By noting that
\beq
\e_s{}_\I \nabla_a \e_\sigma^\I=-\e_\sigma{}_\I \nabla_a \e_s^\I=
\frac{1}{\dot \sigma} (\dot s_a + \dot \theta \sigma_a ),
\label{esDesigma}
\eeq
one finds
\beq
\e_\sigma^\I \Dc_a \Dc^a \p_\I=D^a \sp_a-Y_{(s)},
\eeq
where we have defined
\beq
Y_{(s)}\equiv \frac{1}{\dot \sigma} (\dot s_a + \dot \theta \sigma_a^\perp )s^a\, .
\label{Y_s}
\eeq
Thus, finally, the adiabatic combination of the Klein-Gordon
equations can be written as
\beq
\ddot \sigma + \Theta \dot \sigma + V_{,\sigma}= \nabla^a \sp_a
  -Y_{(s)},
\label{evolsigma}
\eeq
where we have used the property
\beq
D^a\sp_a+a^a\sp_a=\nabla^a\sp_a, \label{grad}
\eeq
which is valid for any covector {\it orthogonal to} $u^a$.

Let us now consider the ``entropic'' combination of the
Klein-Gordon equations, i.e., the contraction of (\ref{evolp})
with $\e_s^\I$. By using
\beq
\e_s^\I \ddot \p_\I=\dot\theta\dot\sigma,
\eeq
and by defining the entropic gradient of the potential,
\beq
V_{, s} \equiv \e_s^\I V_{,\p_\I},
\eeq
one finds
\beq
\dot \sigma\dot\theta +V_{,s} -\e_s^\I \Dc_a \Dc^a \p_\I -a^a
s_a=0.
\eeq
One can rewrite the last two terms by using the identity
\beq
\e_s^\I \Dc_a \Dc^a \p_\I=D^a s_a+Y_{(\sigma)},
\eeq
with
\beq
Y_{(\sigma)}\equiv \frac{1}{\dot \sigma} (\dot s_a + \dot \theta
\sigma^\perp_a )\sp{}^a,
\label{Y_sigma}
\eeq
 and by applying
the property (\ref{grad}) to the  covector $s_a$. One finally gets
\beq
\dot \sigma \dot \theta + V_{,s}= \nabla_a s^a
  + Y_{(\sigma)}. \label{thetadot}
\eeq

To summarize, we have been able to replace the Klein-Gordon
equations for the fields $\phi$ and $\chi$ by two new equations
which embody the evolution along the adiabatic and
entropy directions respectively. The left hand side of these equations has
exactly the same form as the homogeneous equations in a Friedmann
universe. Although our covariant equations (\ref{evolsigma}) and
(\ref{thetadot}) look very similar to the homogeneous equations,
they capture the fully nonlinear dynamics of the scalar fields.
This is manifest in the right hand side of these equations, which
contain nonlinear (quadratic) terms represented by $Y_{(s)}$ and
$Y_{(\sigma)}$, sourcing the adiabatic and entropy equations
respectively. Whereas these equations generalize the {\em
background} evolution equations,  we will go one step further in
the next subsection by deriving fully nonlinear and exact
equations which mimic and generalize the {\em linearized}
equations for the adiabatic and entropy components.

\subsection{Evolution of the adiabatic and entropy covectors}

We now derive evolution equations for the covectors $\sigma_a$ and
$s_a$. More precisely, our purpose is  to find two evolution
equations, which are second order in time (with respect to
the Lie derivative along $u^a$) and  which mimic the equations
obtained in the linear theory for the perturbations
$\delta\sigma$ and $\delta s$ (see \cite{Gordon:2000hv}).

Let us begin with the evolution equation for $\sigma_a$.
Starting from its definition (\ref{tan_ort1}),
 one finds that its  time derivative, i.e., the
Lie derivative with respect to $u^a$, is given by
\beq
\dot\sigma_a=\e_\sigma^\I\nabla_a\dot\p_\I+\dot\theta s_a=\nabla_a\dot\sigma+\dot\theta s_a,
\label{dotsigma}
\eeq
where the last equality is obtained by using $\dot\p_\I=\dot\sigma
\e_{\sigma \I}$. A further time derivative yields
\beq
\ddot \sigma_a =\nabla_a\ddot\sigma+ \ddot \theta s_a
+  \dot \theta \dot s_a.
\eeq
The next step consists in using (\ref{evolsigma})
to eliminate $\ddot\sigma$ in the  above expression.
This gives
\bea
&&\ddot \sigma_a + \Theta \dot \sigma_a +\dot\sigma \nabla_a\Theta +\left(V_{,\sigma
\sigma}+\dot \theta \frac{V_{,s}}{\dot \sigma} \right) \sigma_a -\nabla_a \left(\nabla^c \sp_c\right)
=\left(\dot \theta-\frac{V_{,s}}{\dot \sigma}\right) \dot s_a
  \nonumber \\
&& \qquad  +\left(\ddot\theta - V_{,\sigma s}+ \Theta \dot \theta\right) s_a -
\nabla_aY_{(s)}\ , \label{sigma1}
\eea
where we have used the relation
\beq
\nabla_a V_{,\sigma}= V_{, \sigma \sigma} \sigma_a +V_{,\sigma s} s_a
+\frac{V_{,s}}{\dot \sigma} (\dot s_a + \dot \theta \sigma_a),
\eeq
and introduced the notation
\beq
V_{,\sigma \sigma} \equiv \e_\sigma^\I \e_\sigma^\J V_{,\p_\I
\p_\J }, \qquad V_{,s s} \equiv \e_s^\I \e_s^\J V_{,\p_\I \p_\J },
\qquad V_{,s \sigma} \equiv \e_s^\I \e_\sigma^\J V_{,\p_\I \p_\J
},
\eeq
for the second derivatives of the
potential.

The evolution equation for $\sigma_a$ can be  decomposed
into a longitudinal part, obtained by
contracting (\ref{sigma1}) with  $u^a$, and an orthogonal
part obtained by contraction with
$h_{ab}$. By  using the relation
\beq
\dot V_{,\sigma} = V_{, \sigma \sigma} \dot \sigma + \dot \theta V_{,s},
\eeq
it is not difficult to see that the longitudinal part yields in
fact the time derivative of (\ref{evolsigma}). What is more
interesting is the orthogonal or spatial part which can be written
as
\bea
&&(\ddot \sigma_a)^\perp + \Theta (\dot \sigma_a)^\perp
+\dot\sigma D_a\Theta +\left(V_{,\sigma
\sigma}+\dot \theta \frac{V_{,s}}{\dot \sigma} \right)
\sigma_a^\perp -D_a \left(\nabla^c \sp_c\right)
=\left(\dot \theta-\frac{V_{,s}}{\dot \sigma}\right) \dot s_a
  \nonumber \\
&& \qquad  +\left(\ddot\theta - V_{,\sigma s}+ \Theta
\dot \theta\right) s_a -
D_aY_{(s)}\ . \label{sigma1_spatial}
\eea

Let us now consider the evolution equation for $s_a$. From
Eq.~(\ref{esDesigma}) the time derivative of $s_a$ is given by
\beq
\dot s_a= \dot\sigma \e_s^\I\nabla_a \e_{\sigma\I} -\dot\theta
\sigma_a.
\eeq
Taking another time derivative, one finds
\beq
\ddot s_a = - \ddot \theta \sigma_a -  \dot \theta
\dot \sigma_a+\frac{\ddot\sigma}{\dot\sigma}(\dot s_a + \dot \theta \sigma_a )+\dot\sigma\nabla_a \dot\theta,
\eeq
where we have used (\ref{esDesigma}) and (\ref{angle_1}). We can
now use the entropic equation (\ref{thetadot}) to get rid of
$\nabla_a \dot\theta$. Furthermore, using the relation
\beq
\nabla_a V_{,s}= V_{, s \sigma} \sigma_a +V_{,ss} s_a
-\frac{V_{,\sigma}}{\dot \sigma} (\dot s_a + \dot \theta
\sigma_a),
\eeq
we finally obtain
\bea
&&\ddot s_a - \frac{1}{\dot \sigma}(\ddot \sigma + V_{,\sigma})
\dot s_a +( V_{,ss}-\dot \theta^2 ) s_a
-\nabla_a \left(\nabla_c s^c\right)
=- 2
\dot \theta \dot \sigma_a   \nonumber \\
&& \qquad +\left[\frac{\dot
\theta}{\dot \sigma} (\ddot \sigma+V_{,\sigma})
-\ddot \theta - V_{,\sigma s}\right] \sigma_a+
\nabla_a Y_{(\sigma)}\ .\label{s0}
\eea
As for the adiabatic equation,
 the longitudinal part of this equation,
upon using the relation
\beq
\dot V_{,s}  =  V_{,\sigma s}\dot \sigma - \dot \theta V_{,
\sigma} , \label{dotVs}
\eeq
yields the time derivative of
Eq.~(\ref{thetadot}). The orthogonal part, instead, yields
\bea
&&\ddot s_a - \frac{1}{\dot \sigma}(\ddot \sigma + V_{,\sigma})
\dot s_a +( V_{,ss}-\dot \theta^2 ) s_a
-D_a \left(\nabla_c s^c\right)
=- 2
\dot \theta (\dot \sigma_a)^\perp   \nonumber \\
&& \qquad +\left[\frac{\dot
\theta}{\dot \sigma} (\ddot \sigma+V_{,\sigma})
-\ddot \theta - V_{,\sigma s}\right] \sigma_a^\perp+
D_a Y_{(\sigma)}\ ,\label{s1}
\eea
where we have used  the property that the covectors $\dot s_a$ and
$\ddot s_a$ are purely spatial, i.e., that $(\dot s_a)^\perp =
\dot s_a $ and $(\ddot s_a)^\perp = \ddot s_a $.

Starting from the fully nonlinear Klein-Gordon equations, we have
thus managed to obtain a system of two coupled equations
(\ref{sigma1_spatial}) and (\ref{s1}), which govern the evolution
of our nonlinear adiabatic and entropy components. These equations
are one among the main results of the present work. Remarkably, they
are rather simple and they look very similar to their linearized
counterparts. As we will see later, immediately deduced from
these equations are the already known evolution  equations for the
{\em linear} adiabatic and entropy components. Furthermore, since
our equations are exact, they can be used  to go beyond the linear
order,  up to second or higher orders. To illustrate this, we will
show explicitly in Sec.~\ref{sec:secondorder} how to derive
from these equations, in a
systematic way,  the evolution of the second-order adiabatic and
entropy components.

%%%%%%%%%%%%%%%%%%%%%%%%%%%%%%%%%%%%%%%%%%%%%%%%%%%%%%%%%%%
\subsection{Generalized covariant perturbations}
%%%%%%%%%%%%%%%%%%%%%%%%%%%%%%%%%%%%%%%%%%%%%%%%%%%%%%%%%%%

In this subsection, we will be first interested in the covariant
generalization of the comoving energy density and curvature
perturbations in the context of a two-field system. We will then
consider the generalization of the curvature perturbation on
uniform energy density hypersurfaces.

Let us  introduce the covector
\beq
\epsilon_a\equiv\Dc_a\rho- \frac{\dot \rho}{\dot \sigma}\sp_a,
\label{epsilon}
\eeq
which can be interpreted as a covariant generalization of the {\em comoving
energy density} perturbation. In order to obtain  the explicit expression of
$\epsilon_a$ in terms of $\sigma_a$ and $s_a$, let us rewrite the components (\ref{rhotot}-\ref{as})
of the energy-momentum tensor in the form
\bea
\rho \tackl = \tackr \frac{1}{2} \left(   \dot \sigma^2 + \Pi  \right) +V ,
\label{EMTrho} \\
P \tackl = \tackr \frac{1}{2} \left(   \dot \sigma^2 -\frac13 \Pi  \right) -V ,
\label{EMTP} \\
q_a \tackl = \tackr -\dot\sigma \sp_a, \label{qa} \\
\pi_{ab} \tackl = \tackr    \Pi_{ab} -\frac13 h_{ab} \Pi, \label{pi}
\eea
where we have defined
\beq
\Pi_{ab} \equiv \sp_a \sp_b +s_a s_b,  \qquad \Pi \equiv \sp_c
{\sp}^c +s_c s^c\, .
\eeq
Substituting the expression for $\rho$ into the definition
(\ref{epsilon}), one gets
\beq
\epsilon_a = \dot \sigma (\dot \sigma_a)^\perp - \ddot
\sigma\sigma_a^\perp +  (V_{, s} -\dot \theta
\dot \sigma) s_a +\frac{1}{2} \left(\Dc_a \Pi -\frac{\dot
\Pi}{\dot \sigma} \sp_a \right) , \label{epsilon_inter}
\eeq
where we have used the relation
\beq
 D_a \dot \sigma =  \dot \sigma_a +\ddot\sigma u_a -\dot \theta  s_a,
\eeq
which is a consequence of (\ref{dotsigma}).

As noticed above, the longitudinal projection of Eq.~(\ref{s0})
yields the time derivative of (\ref{thetadot}), which reads
\beq
\frac{\dot \theta}{\dot \sigma} V_{,\sigma}-\ddot \theta -
V_{,\sigma s}=\frac{\dot \theta}{\dot \sigma} \ddot
\sigma-\frac{1}{\dot\sigma}\Tdot{\left(\Dc_cs^c+Y_{(\sigma)}\right)}.
\label{relation_ddtheta}
\eeq
This  relation, together with Eqs.~(\ref{evolsigma}) and
(\ref{epsilon_inter}),  can be employed to rewrite
Eq.~(\ref{s1}) in the form
\bea
&&\ddot s_a +\left[\Theta-\frac{1}{\dot \sigma} \left(\nabla^c
\sp_c -Y_{(s)} \right) \right]\dot s_a + \left( V_{,ss}+\dot
\theta^2-2 \dot \theta \frac{V_{,s}}{\dot \sigma} \right) s_a -D_a
\left(\nabla_c s^c\right)=
\nonumber \\
&& \qquad - 2 \frac{\dot \theta}{\dot \sigma}  \epsilon_a
+\frac{\dot \theta}{\dot \sigma} \left( \Dc_a \Pi -
\frac{\dot \Pi}{\dot \sigma} \sp_a \right)
 -
\frac{1}{\dot\sigma} \Tdot{\left(\Dc_cs^c+Y_{(\sigma)}\right)}\sigma_a^\perp
+D_a Y_{(\sigma)}.  \label{s3}
\eea
As in the linear theory (see \cite{Gordon:2000hv}),  this gives us
an alternative expression for the  evolution equation for  $s_a$,
in which  the comoving energy density perturbation appears
explicitly on the right hand side. This expression will be useful
in Sec.~\ref{sec:ls} when discussing the large-scale evolution of
$s_a$.

Together with the comoving energy density, 
the comoving curvature perturbation can generalized. For the general case of several
scalar fields,  this is done by defining the comoving
integrated expansion perturbation,
\beq
\R_a \equiv   - D_a \alpha  -  \frac{\dot \alpha}{(\dot \varphi_J
\dot \varphi^J)} q_a. \label{R_N}
\eeq
The definition of the Sasaki-Mukhanov covector given for a single
scalar field in Eq.~(\ref{Q_single}) can then be extended to the
case of several fields, by defining for each field
\beq
Q^I_a \equiv D_a \varphi^I - \frac{\dot \varphi^I}{\dot \alpha}
D_a \alpha .
\eeq
Thus, the comoving covector $\R_a$ can also be written as
\beq
\R_a = \frac{\dot \alpha}{(\dot \varphi_J \dot \varphi^J)}\  \dot
\varphi_I Q^I_a .
\eeq
In the {\em two}-field case, the definition (\ref{R_N}) reduces,
using (\ref{qa}),  to
\beq
\R_a \equiv - D_a \alpha + \frac{\dot \alpha}{\dot \sigma}
\sigma_a^\perp. \label{R_two}
\eeq
Furthermore, one can generalize the Sasaki-Mukhanov variable to
the {\em adiabatic} covector by defining
\beq
Q_a \equiv\e_{\sigma \I} Q^I_a = \sigma_a^\perp - \frac{\dot \sigma}{\dot \alpha} D_a
\alpha . \label{Q_two}
\eeq

Instead of giving the evolution equation of $\R_a$ we will resort
to a fluid description by considering the covariant
generalization of the uniform density curvature perturbation,
i.e., the integrated expansion perturbation on
uniform density hypersurfaces  $\zeta_a$, defined
in Eq.~(\ref{zeta}). In the two-field case, these two quantities
are related by
\beq
\zeta_a + \R_a = - \frac{\dot \alpha}{\dot \rho} \epsilon_a.
\label{zeta_R_two}
\eeq

In contrast with the case of a single scalar field, which can
always be described as a perfect fluid, the total energy-momentum
for two or more scalar fields will in general correspond to that
of a dissipative fluid and the nonlinear formalism developed in
\cite{Langlois:2006iq} will thus be useful in this case.

The adiabatic Klein-Gordon equation (\ref{evolsigma}) can be rewritten as
a continuity equation for the total energy density (\ref{EMTrho}) and pressure
(\ref{EMTP}), which reads
\beq
\label{conservation}
\dot \rho + \Theta (\rho + P) = {\cal D},
\eeq
with the ``dissipative'' term
\beq
\D =  \dot \sigma \left( \nabla^a \sigma_a^\perp - Y_{(s)} \right)
+ \frac{1}{3}
\Theta \Pi + \frac{1}{2} \dot \Pi.
\eeq

In \cite{Langlois:2006iq} it was shown that the evolution equation
for $\zeta_a$ for a dissipative fluid, which generalizes (\ref{zeta_evolution_single}),  is
given by
\beq
\dot \zeta_a = \frac{\Theta^2}{3\dot \rho} \left(\Gamma_a +
\Sigma_a \right),
\label{zeta_evolution}
\eeq
where the second source term on the right hand side, due to the dissipative nature
of the fluid,
is defined in terms of ${\cal D}$ as
\beq
\Sigma_a \equiv - \frac{1}{\Theta} \left(D_a \D - \frac{\dot \D}{\dot
\rho} D_a \rho\right) + \frac{\D}{ \Theta^2} \left(D_a \Theta -
\frac{\dot \Theta}{\dot \rho} D_a \rho \right) . \label{Sigma}
\label{Sigma_def}
\eeq
We have called this term the {\em
dissipative} nonadiabatic pressure perturbation because it has the
same form as $\Gamma_a$, provided  one replaces the pressure
$P$ by a dissipative pressure $-{\cal D}/\Theta$ defined in terms of
dissipative fluid quantities.

We now concentrate on the covector $\Gamma_a$ and rewrite it by
taking into account the fact that our effective fluid consists of
two scalar fields. In this case, the pressure and energy density
are related by
\beq
P= \rho - 2V - \frac{2}{3}  \Pi,\label{Ptorho}
\eeq
as can be inferred from Eqs.~(\ref{EMTrho}--\ref{EMTP}). Inserting
this relation into the definition of $\Gamma_a$,
Eq.~(\ref{Gamma_def}), one gets
\beq
\Gamma_a = \left( 1-\frac{\dot P}{\dot \rho} \right) D_a \rho -2
D_a V - \frac{2}{3} D_a \Pi.
\eeq
Using  the time
derivative of (\ref{Ptorho})
and the relation
\beq
D_a V= V_{,\sigma} \sp_a + V_{,s} s_a,
\eeq
one finally obtains
\beq
\Gamma_a = 2 \frac{\dot \sigma}{\dot \rho} V_{,\sigma}  \epsilon_a
-2 V_{,s} s_a  -\frac{2}{3} \left(D_a \Pi -\frac{\dot \Pi}{\dot
\rho} D_a \rho \right),
\label{Gamma_1}
\eeq
where we have introduced  $\epsilon_a$ defined in
Eq.~(\ref{epsilon}). The above equation expresses the
nonlinear nonadiabatic pressure perturbation in the two-field
case, which sources Eq.~(\ref{zeta_evolution}). It will be useful
in the next section, where we will consider both the linear and
super-Hubble approximations of our evolution equations.

The above equation expresses the nonlinear nonadiabatic pressure
perturbation in the two-field case. The evolution of $\zeta_a$,
governed by Eq.~(\ref{zeta_evolution}), is thus sourced by a
rather complicated term obtained by summing $\Gamma_a$ in
Eq.~(\ref{Gamma_1}) and $\Sigma_a$ in Eq.~(\ref{Sigma_def}). As we
will see in the next section, this term simplifies considerably
when either the linear or the super-Hubble approximations are
taken.

%%%%%%%%%%%%%%%%%%%%%%%%%%%%%%%%%%%
%%%%%%%%%%%%%%%%%%%%%%%%%%%%%%%%%%%
\section{Approximate equations}
\label{sec:approximate}
%%%%%%%%%%%%%%%%%%%%%%%%%%%%%%%%%%%
%%%%%%%%%%%%%%%%%%%%%%%%%%%%%%%%%%%

 In this section we study the evolution equations of the
two-field system under two types of approximations: the linear
limit and the limit where we neglect higher orders in spatial
gradients. In an expanding FLRW universe, the latter corresponds
to the {\em large scale} limit. In the rest of the paper we will
use the symbol $\simeq $ to denote an equality at the linear level,
and the symbol $\approx $ to denote an equality valid only on large
scales.

\subsection{Homogeneous and linearized equations}

In many cosmological applications, since our Universe appears to
be close to a FLRW universe on large scales, it is sufficient to
restrict oneself to  the {\it linearized} version of the evolution
equations. Before working within some generic coordinate system in
the next section, we first consider this linearization procedure
directly at the level of  the covariant equations, as in
\cite{Ellis:1989jt}.

In a strictly FLRW universe, all the spatial gradients vanish and
therefore
\beq
\sigma_a^\perp=0, \qquad s_a=0.\qquad ({\rm FLRW})
\eeq
Consequently, the scalar quantities $Y_{(s)}$ and $Y_{(\sigma)}$,
defined respectively in (\ref{Y_s}) and (\ref{Y_sigma}),  vanish
and the evolution equations for $\sigma$ and $\theta$,
respectively (\ref{evolsigma}) and (\ref{thetadot}), reduce to
\beq
\ddot \sigma + 3H \dot \sigma + V_{,\sigma}= 0,
\label{evolsigma_RW} \qquad ({\rm FLRW})
\eeq
and
\beq
\dot \sigma \dot \theta + V_{,s}= 0, \label{thetadot_RW} \qquad ({\rm FLRW})
\eeq
where we have introduced the Hubble parameter $H=\Theta/ 3$. Not surprisingly,
the above equations correspond exactly to the homogeneous equations
given in \cite{Gordon:2000hv}. Furthermore,
in  a FLRW universe, all the terms in  Eq.~(\ref{s1}) for $s_a$ vanish.
This is the same for Eq.~(\ref{sigma1_spatial})
for $\sigma_a$ projected orthogonally to
$u_a$. Although its longitudinal component does not vanish it is simply the
time derivative of (\ref{evolsigma_RW}).

At linearized order, we treat  the covectors $\sigma_a^\perp$ and
$s_a$, which vanish at zeroth order, as {\it first-order}
quantities. Similarly, their derivatives $\dot s_a$, $\ddot s_a$,
$(\dot \sigma_a)^\perp$ and $(\ddot\sigma_a)^\perp$ are
first-order quantities. Therefore the linearized version of the
evolution equations are simply obtained by keeping only the
homogeneous terms in the coefficients multiplying the spatial
projection of $\sigma_a$, $s_a$ and their derivatives. The
linearized versions of (\ref{sigma1_spatial}) and (\ref{s3}) are
thus,  respectively,
\beq
(\ddot \sigma_a)^\perp + 3H (\dot \sigma_a)^\perp + \dot \sigma D_a \Theta
+\left(V_{,\sigma
\sigma}-\dot \theta^2 \right) \sigma_a^\perp -D_a \left(D^c \sp_c\right)
\simeq
2 \Tdot{\left( \dot \theta s_a\right)}
  \nonumber -2 \frac{V_{,\sigma}}{\dot \sigma}  \dot \theta
s_a , \label{sigma1_spatial_lin}
\eeq
and
\beq
\ddot s_a +3H \dot s_a +\left( V_{,ss}+3\dot \theta^2 \right) s_a
-D_a \left(D_c s^c\right) \simeq- 2 \frac{\dot \theta}{\dot \sigma}
\epsilon_a .\label{s1_lin}
\eeq
Note that the terms involving $Y_{(s)}$, $Y_{(\sigma)}$ and $\Pi$
have disappeared, since these scalars are {\it quadratic} in
first-order quantities. We have also replaced $\nabla^c \sp_c$ by
$D^c \sp_c$, as well as $\nabla^cs_c$ by $D^cs_c$, since their
difference is quadratic in first-order quantities, according to
(\ref{grad}). Indeed,  the acceleration vector $a^b$, which
vanishes at zeroth order, is considered as a first-order quantity.
Finally, we have used Eqs.~(\ref{evolsigma_RW}--\ref{thetadot_RW})
(together with the time derivative of the latter equation), to
rewrite the right hand side of the linearized equations in the
above form.

One can also linearize the evolution equation for $\zeta_a$.
As discussed above, the terms containing
$Y_{(s)}$ and $\Pi$ can be neglected.
The dissipative term ${\cal D}$  thus reduces to

\beq
{\cal D}\simeq \dot \sigma D^a \sigma_a^\perp,
\eeq
while the expression (\ref{Gamma_1}) for $\Gamma_a$ becomes simply
\beq
\Gamma_a \simeq 2 \frac{\dot \sigma}{\dot \rho} V_{,\sigma}  \epsilon_a
-2 V_{,s} s_a.
\eeq
Therefore, the evolution equation for $\zeta_a$, Eq.~(\ref{zeta_evolution}), can
be written, at linear order, as
\beq
\dot \zeta_a \simeq  \frac{2}{3\dot \sigma^3}   V_{,\sigma} \epsilon_a
+ \frac{2H}{\dot \sigma^2} V_{,s} s_a + \frac{1}{3 \dot \sigma}
D_a (D^b \sigma_b^\perp). \label{zeta_evolution_lin}
\eeq

As expected, the linearized equations
(\ref{sigma1_spatial_lin}--\ref{s1_lin}) and
(\ref{zeta_evolution_lin}) are equivalent to the linear equations
obtained in the coordinate based approach in \cite{Gordon:2000hv}.
This will be even more explicit in Sec.~\ref{sec:linear}  where we
introduce a generic coordinate system and compute explicitly the
components of our various tensors and equations.

%%%%%%%%%%%%%%%%%%%%%%%%%%%%%%%%%%%%%%%%%%%%%
\subsection{Expansion in spatial gradients}
\label{sec:ls}
%%%%%%%%%%%%%%%%%%%%%%%%%%%%%%%%%%%%%%%%%%%%%
Apart from the linearization procedure, there is another
approximation in  the cosmological context which applies to
describe the Universe on very large scales, even  beyond the
present Hubble radius where in principle  it could strongly
deviate from a FLRW universe. This approximation is based on an
expansion in spatial gradients, which are small for scales larger
than the local Hubble radius
\cite{Salopek:1990jq,Comer:1994np,Deruelle:1994iz}.

In this perspective one sees, from their definition (\ref{perp}),
that $\sigma_a^\perp$ and $s_a$ are first-order quantities with
respect to spatial gradients because they are linear combinations
of spatial gradients. The scalars $Y_{(s)}$ and $Y_{(\sigma)}$ are
second order with respect to spatial gradients since they are
quadratic in $\sigma_a^\perp$ and $s_a$ (or their time
derivatives). Hence, the right hand side of Eq.~(\ref{evolsigma})
and of Eq.~(\ref{thetadot}) can be neglected on large scales, so
that these two equations become, in the large-scale limit,
\beq
\ddot \sigma + \Theta \dot \sigma + V_{,\sigma} \approx 0,
\label{sigma_evol_ls}
\eeq
and
\beq
\dot \theta  \approx  - \frac{V_{,s}}{\dot \sigma }.
\label{thetadot_ls}
\eeq
Although they look very similar to the homogeneous equations
(\ref{evolsigma_RW}) and (\ref{thetadot_RW}), these equations are
fully inhomogeneous and encode the evolution of nonlinearities on
large scales. This limit illustrates the separate universe picture
\cite{Sasaki:1995aw,Starobinsky} where the inhomogeneous universe
can be described, on large scales, as juxtaposed Friedmann
homogeneous universes.

If, so far, the order in spatial gradients seems to coincide with
the perturbative classification of the previous subsection, it
differs however for the term
\beq
\nabla^c \sp_c = D^c \sp_c+a^c\sp_c ,
\eeq
which is first order perturbatively but second order in spatial
gradients, at least for the first term on the right hand side,
since $\sp_c$ is already first order in spatial gradients.
 For the second term,
we will assume that $u^a$ can be chosen so that $a^c$ is at least
first order in spatial gradients. We will show this explicitly in
the next section by working in a coordinate system.

With these prescriptions, the evolution equation of
$\sigma_a^\perp$ and $s_a$ obtained at lowest order in  spatial
gradients become
\beq
(\ddot \sigma_a)^\perp + \Theta (\dot \sigma_a)^\perp +\dot\sigma
D_a\Theta +\left(V_{,\sigma \sigma}-\dot \theta^2 \right)
\sigma_a^\perp  \approx 2 \Tdot{\left(\dot \theta s_a \right)} -2
\dot \theta \frac{V_{,\sigma}}{\dot \sigma}
 s_a
, \label{sigma2}
\eeq
and
\beq \ddot s_a +\Theta \dot s_a + \left( V_{,ss}+3 \dot
\theta^2 \right) s_a  \approx - 2 \frac{\dot \theta}{\dot \sigma}
\epsilon_a , \label{s4}
\eeq
where we have dropped the terms containing $\Pi$ in (\ref{s3}), which are two
orders higher than $s_a$ in  spatial gradients.

We now expand the evolution equation for $\zeta_a$,
Eq.~(\ref{zeta_evolution}), by neglecting higher-order spatial
gradients in the two terms on the right hand side of this
equation. The nonadiabatic pressure perturbation becomes
\beq
\Gamma_a \approx 2 \frac{ \dot \sigma}{\dot \rho} V_{,\sigma}
\epsilon_a -2 V_{,s} s_a , \label{Gamma_ls}
\eeq
while the dissipative nonadiabatic pressure perturbation
$\Sigma_a$ can be completely dropped, since the dissipative term
$\cal D$ is at least second order in the spatial gradients and
thus $\Sigma_a$ is third order in the spatial gradients. Equation
(\ref{zeta_evolution}) therefore becomes, on large scales,
\beq
\dot \zeta_a \approx \frac{2}{3 \dot \sigma^3}
V_{,\sigma}  \epsilon_a
+\frac{2 \Theta}{ 3 \dot \sigma^2}
V_{,s} s_a  . \label{zeta_ls}
\eeq
Note that the lowest order limit in  spatial gradients of the
evolution equations of $\sigma_a^\perp$, $s_a$ and $\zeta_a$,
respectively  Eqs.~(\ref{sigma2}--\ref{s4}) and (\ref{zeta_ls}),
are similar to their linear counterparts  except for the terms
$D_a (D^c \sigma_c^\perp)$ and $D_a (D^c s_c)$, which are third
order in  spatial gradients  and therefore negligible in the
spatial gradient expansion. This is because in these equations
the  terms that are higher than linear order in the perturbative
expansion turn out to be  also higher than first order in the
spatial gradient expansion.

We now concentrate our attention on the comoving energy density
perturbation, $\epsilon_a$, defined  in Eq.~(\ref{epsilon}). Until
now we have only made use of the Klein-Gordon equations of the
scalar fields. However, in order to study the behavior of the
comoving energy density, we will now make use of the Einstein
equations, in particular of the so-called {\em constraint
equations}. The projection of the Einstein equations along $u^a$
yields the  {\em energy constraint} in a covariant form,
\beq
u^a G_{ab}u^b=8\pi G \rho.
\eeq
If one assumes that the vector field $u^a$ is hypersurface
orthogonal, it is possible to use the Gauss-Codacci equations
(see e.g., \cite{wald}) and
the decomposition
\beq
D_b u_a=\sigma_{ab}+{1\over 3}\Theta h_{ab},
\label{decomposition_spatiale}
\eeq
which is the spatially projected version of (\ref{decomposition})
(with $\omega_{ab}=0$ since
$u^a$ is here hypersurface orthogonal),
in order to  rewrite the energy constraint as
\beq
\frac{1}{2}\left({}^{(3)} \! R+\frac{2}{3}\Theta^2-\sigma_{ab}
\sigma^{ab}\right)=8\pi G\rho, \label{e_c}
\eeq
where ${}^{(3)} \! R$ is the intrinsic  Ricci
scalar of the space-like hypersurfaces orthogonal to $u^a$.

The mixed projection of Einstein's equations yields the covariant
{\em momentum constraint}
\beq
u^b G_{bc}h^c_a= 8\pi G q_a,
\eeq
which can be rewritten, via Gauss-Codacci relations and
Eq.~(\ref{decomposition_spatiale}), as
\beq
D_b\sigma_a^{\ b}-\frac{1}{3} D_a\Theta=8\pi G q_a. \label{m_c}
\eeq
By combining the energy and momentum constraints, one obtains the
{\it nonlinear covariant} version of the generalized Poisson
equation, which in the linear theory relates the comoving energy
density to the Bardeen's potential defined from the curvature
perturbation. Here one finds
\beq
\frac{1}{2}D_a\left({}^{(3)} \! R -\sigma_{bc}\sigma^{bc}
\right)+\Theta D_b\sigma^b_a=8\pi G {\tilde\epsilon}_a,
\label{poisson2}
\eeq
where we have introduced on the right hand side
\beq
{\tilde\epsilon}_a \equiv D_a\rho-\Theta
q_a=D_a\rho+\Theta\dot\sigma\sigma_a^\perp. \label{epsilon_ls_tilde}
\eeq
This quantity can be seen as an alternative generalization of the
comoving energy density perturbation, since, in the linear limit,
it is equivalent to $\epsilon_a$ defined in (\ref{epsilon}). In
the fully nonlinear case, the two quantities differ and, using
(\ref{conservation}), one finds
\beq
{\tilde\epsilon}_a-\epsilon_a=\frac{1}{\dot\sigma}\left(\D-\frac{1}{3}\Theta\Pi\right)\sigma_a^\perp.
\eeq
This difference becomes however negligible on large scales,
\beq
\epsilon_a \approx \tilde \epsilon_a.
\label{eps_eq_eps_tilde}
\eeq

Now, the left hand side of Eq.~(\ref{poisson2}) contains the
projected gradient of the Ricci scalar, $D_a {}^{(3)} \! R$. From
its definition in terms of derivatives of the metric \cite{wald},
it can be shown that, for a perturbed FLRW universe, this term is
of third order in the spatial gradients. Equation (\ref{poisson2})
can thus be used to show that $\epsilon_a$ in
Eqs.~(\ref{zeta_R_two}), (\ref{s4}) and (\ref{zeta_ls}) can be
neglected on large scales, if the shear can also be neglected in
this limit. Indeed, on large scales, the shear rapidly decreases
in an expanding perturbed FLRW universe. Thus, in this limit the
comoving and uniform density integrated expansion
perturbations $\zeta_a$ and $\R_a$ {\em coincide} (up
to a sign),
\beq
\zeta_a + \R_a \approx 0. \label{zeta_R_ls}
\eeq
Furthermore, one can rewrite Eqs.~(\ref{s4}) and (\ref{zeta_ls})
as a closed coupled system of equations, describing the
large-scale nonlinear evolution of adiabatic and entropy
perturbations,
\beq
\ddot s_a +\Theta \dot s_a + \left( V_{,ss}+3 \dot \theta^2
\right) s_a \approx 0, \label{s_evol_ls}
\eeq
and \beq \dot \zeta_a \approx - \frac{2}{3}  \frac{\Theta^2}{ \dot
\rho} V_{,s} s_a \label{zeta_evol_ls}.
\eeq
These two equations are analogous to the equations derived for
nonlinear perturbations by Rigopoulos et
al.~\cite{Rigopoulos:2005} in the long wavelength approximation.

%%%%%%%%%%%%%%%%%%%%%%%%%%%%%%%%%%%%%%%%%%%%%%%%%%%%%%%%%%%%%%%%%%%%%%%%%%%%%%%%%%%%%%%%%%%%%%
\section{Linear perturbations}
\label{sec:linear}
%%%%%%%%%%%%%%%%%%%%%%%%%%%%%%%%%%%%%%%%%%%%%%%%%%%%%%%%%%%%%%%%%%%%%%%%%%%%%%%%%%%%%%%%%%%%%%

We now relate  our covariant approach with the more familiar
coordinate based formalism. We first examine the linear
perturbations in the present section and we consider
second-order perturbations in the next one.

Let us thus  introduce generic  coordinates $x^\mu=\{t, x^i\}$ to
describe an  almost-FLRW spacetime. Here a  prime will denote a partial
derivative with respect to the cosmic  time $t$, i.e. ${}' \equiv
\partial / \partial t$, since the dot has been reserved to 
denote the Lie derivative
with respect to $u^a$.

The background spacetime is a FLRW
spacetime, endowed with the metric
\beq
ds^2={\bar g}_{\mu\nu}dx^\mu dx^\nu=-dt^2+
a(t)^2\gamma_{ij}dx^i dx^j.
\eeq
At linear order, the  spacetime geometry  is described by the perturbed metric
\beq
ds^2=\left({\bar g}_{\mu\nu}+\d g_{\mu\nu}\right)dx^\mu dx^\nu,
\eeq
where the components of the metric perturbations can  be written
as
\beq
\delta g_{00}=-2A, \quad \delta g_{0i}=a B_i, \quad \delta
g_{ij}=a^2 H_{ij}.
\eeq
As usual we  decompose $B_i$ and $H_{ij}$ in the forms
\bea
B_i \tackl = \tackr \vec \nabla_i B + B^V_i, \\
\label{H} H_{ij} \tackl = \tackr 
-2\psi \gamma_{ij}+2 \vec\nabla_i\vec\nabla_j E+2
\vec \nabla_{(i}E^V_{j)}+2 { E}^T_{ij},
\eea
where $B_i^V$ and $E^V_i$ are transverse, i.e., 
$\vec \nabla_i B^V{}^i=0=\vec \nabla_iE^V{}^i$, and 
${E}^T_{ij}$ is transverse and traceless, i.e., $\vec
\nabla_i{E}^T{}^{ij}=0$ and $\gamma^{ij}{E}^T_{ij}=0$. Here $\vec\nabla_i$
denotes the three-dimensional covariant derivative with respect 
to the homogeneous
spatial metric $\gamma_{ij}$ (which is also used to lower or raise
the spatial indices). The matter fields are similarly decomposed into a background and a perturbed
part,
\beq
\p_\I(t,x^i)={\bar\p_\I}(t)+\d \p_\I(t,x^i).
\eeq

We  now need to specify the components of the unit vector $u^a$, which defines the time
derivation in our covariant approach. At zeroth order, it is of course natural to
take it orthogonal to the homogeneous slices. At first order
we choose, for simplicity, $u^\mu$ such that $u_i=0$. This implies that, up to first order, the components
of $u^\mu$ are given by
\beq
\label{components_u}
u^\mu=\{1-A, -B^i/a\} ,
\eeq
 and those of the ``acceleration'' vector are given by
\beq
a^\mu=\{0, \vec\nabla^i A /a^2\}\ .
\eeq
This confirms that  $a^\mu$ can be considered
as first order in spatial gradients, in agreement with our assumption of the previous section.

Since our formalism relies on many covectors, it is useful to first consider a generic covector $X_a$
and work out the  components of its time derivative $\dot X_a$. To make the explicit calculation, it
is convenient to replace in the definition of the Lie derivative (\ref{Lie})
the covariant derivatives by partial derivatives and  write
\beq
\dot X_a= u^b \partial_b X_a + X_b \partial_a u^b.
\label{Lie2}
\eeq
At zeroth order, the
components of $\dot X_a$ are simply
\beq
{\bar {\dot X}}_\mu=\{{\bar X}_0', 0,0,0\},
\eeq
assuming that the spatial components $\bar{X}_i$ vanish so as to respect the symmetries of the geometry.
At first order, we get from (\ref{Lie2}) and (\ref{components_u})
\beq
\label{delta_dot_X}
\delta(\dot X_0)=\delta X_0'-(\bar{X}_0 A)', \qquad
\delta(\dot X_i)=\delta X_i'-\bar{X}_0 \partial_i A.
\eeq

These results can be applied to the particular case of the covector $\phi_a\equiv \nabla_a\phi$, which we introduced
earlier. Its components are, by definition,
\beq
\bar{\phi}_0=\phib',\qquad \bar{\phi}_i=0,
\eeq
at zeroth order and
\beq
\delta\phi_0={\phif}', \qquad \delta \phi_i = \partial_i \phif,
\qquad
\eeq
at first order. Specializing the relations (\ref{delta_dot_X}) to
$X_a=\phi_a$, one gets
\beq
\delta(\dot\phi_0)={\phif}''-\phib'A'-\phib''A, \qquad
\delta(\dot\phi_i)=\partial_i\left({\phif}'-\phib'A\right).
\eeq
Applying once more (\ref{delta_dot_X}), now with $X_a=\dot\phi_a$, one finds
\beq
\delta(\ddot\phi_i)=\partial_i\left({\phif}''-\phib'A'-2\phib''A\right).
\eeq
We will not need
the time component of $\ddot\phi_a$, which involves third order derivatives with
respect to cosmic time.

By using the above results one finds that, at linear order,
the spatial components of Eq.~(\ref{eq_phi_a})
for a single scalar field correspond to the spatial gradient of the following scalar equation:
\beq
{\phif}''+3H{\phif}'+ \bar
V_{,\phi\phi}\phif-\frac{1}{a^2}\vec{\nabla}^2\phif=-2 \bar
V_{,\phi} A+{\phib}'\left[A'+3\psi'
-{\vec\nabla}^2(E'-B/a)\right],
\eeq
where we have used
\beq
\bar\Theta=3H, \qquad
\delta\Theta=-3HA-3\psi'+{\vec\nabla}^2(E'-B/a),
\label{delta_Theta}
\eeq
as well as  the background equation
\beq
\phib''+3H\phib'+ \bar V_{,\phi}=0 \, .
\eeq
 In
the case of several scalar fields, the same equation applies for
each $\varphi_I$ with the replacement of
$\bar V_{,\phi\phi}\phif$ by $\bar V_{,\varphi_I \varphi_J}\delta\varphi^J$.

Let us now consider, for the special case of two scalar fields,
the adiabatic and entropic covectors $\sigma_a$ and $s_a$, which
we have introduced earlier.  Note that with the choice of
four-velocity (\ref{components_u}) $\sigma_i^\perp=\sigma_i$. The
background equations of motion can be deduced immediately from
Eqs. (\ref{evolsigma}) and (\ref{thetadot}) and read
\beq
{\sib}'' + 3H  {\sib}' +\bar V_{,\sigma}= 0, \label{evolsigma_b}
\eeq
\beq
\sib'  \thetab' + \bar V_{,s}= 0, \label{thetadot_b}
\eeq
with
\beq
\sib_0={\bar{\sigma}}'\equiv \sqrt{{\phib}^{\,'2}+{\chib}^{\,'2}}.
\label{sigma_bar}
\eeq
From its definition,  Eq.~(\ref{tan_ort1}), one finds that
the {\it spatial} components of $\sigma_a$  at linear order can be
expressed as
\beq
\delta\sigma_i=\frac{{\phib}'}{{\sib}'}\partial_i\phif+\frac{{\chib}'}{{\sib}'}\partial_i\chif
= \partial_i\delta\sigma,
\label{dsigma}
\eeq
{\it with } the notation
\beq
\sif\equiv
\frac{{\phib}'}{{\sib}'}\phif+\frac{{\chib}'}{{\sib}'}\chif,
\label{sif}
\eeq
which coincides with the definition of
\cite{Gordon:2000hv}.
By using Eq.~(\ref{delta_dot_X}) with (\ref{sigma_bar}-\ref{dsigma})
one finds that the spatial components of the first and second time derivatives are given by
\beq
\label{sigma_comp}
\delta(\dot\sigma_i)=\partial_i\left({\sif}'-\sib'A\right), \qquad
\delta(\ddot\sigma_i)=\partial_i\left({\sif}''-\sib'A'-2\sib''A\right).
\eeq

The same procedure for $s_a$ gives
\beq
\delta s_i=\partial_i \sf, \qquad
\sf\equiv\frac{{\phib}'}{{\sib}'}\chif-\frac{{\chib}'}{{\sib}'}\phif,
\eeq
which also coincides with the notation of \cite{Gordon:2000hv}. Since $s_a$, in contrast to $\sigma_a$,
has no longitudinal component, $\sb_0=0$ and the spatial components of $\dot s_a$ and $\ddot s_a$
are simply
\beq
\label{s_comp} \delta(\dot s_i)=\partial_i{\sf}', \qquad
\delta(\ddot s_i)=\partial_i{\sf}''.
\eeq

Plugging the explicit components (\ref{sigma_comp}) and
(\ref{s_comp}) into the linearized equations for $\sigma_a$ and
$s_a$, given by (\ref{sigma1_spatial_lin}) and (\ref{s1_lin})
respectively, one obtains easily  the linearized equations for
$\delta\sigma$ and $\delta s$. These read, respectively,
 \bea
&& \sif''+3H \sif' + \left( \bar V_{,\sigma \sigma}-\thetab'{}^2
\right) \sif -\frac{1}{a^2} \vec\nabla^2 \sif = 2 ( \thetab' \sf
)' -2\frac{\bar V_{,\sigma} }{\sib'} \thetab' \sf \cr &&  \quad -2
\bar V_{,\sigma}A+\sib'\left[A'+3\psi'-\vec
\nabla^2(E'-B/a)\right] \, , \label{sigma_evol_1}
\eea
and
\beq
\sf'' +3H\sf' +( \bar V_{,ss}+3 \thetab'{}^2 )
\sf-\frac{1}{a^2}\vec\nabla^2\sf =- 2 \frac{\thetab'}{\bar
\sigma'} \delta \epsilon\, .
\label{s_evol_1}
\eeq
In the  latter equation we have introduced  the first-order
comoving energy density perturbation $\delta\epsilon$, defined by
\beq
\delta \epsilon_i = \partial_i \delta \epsilon, \qquad \delta
\epsilon \equiv \delta \rho - \frac{\rhob'}{\sib'} \sif ,
\eeq
which follows from  the definition (\ref{epsilon}) of $\epsilon_a$.
Using
\beq
\rhob = \frac{1}{2} \sib'^2 + \bar V, \qquad \rhof =  \sib'
\sif' - A \sib'^2 + \bar V_{,\sigma} \sif + 2\bar V_{,s} \sf,
\eeq
one sees that $\delta \epsilon$ can be expressed as
\beq
\delta \epsilon = \sib' \sif'-\sib^{\prime 2}A-\sib''\sif + 2
\bar V_{,s} \sf \, .
\label{delta_epsilon_1}
\eeq

Moreover, linearizing the spatial components of the energy
constraint (\ref{e_c}) yields
\beq
3H\left(\psi'+HA\right)-\frac{1}{a^2}{\vec
\nabla}^2\left[\psi+H(a^2E'-aB)\right]=-4\pi G\, \delta\rho,
\label{energy_constraint_1}
\eeq
while the momentum constraint (\ref{m_c}) gives, since $\delta q_i=-\sib'\partial_i\sif$,
\beq
\psi'+HA=4\pi G\sib'\sif.
\label{momentum_constraint_1}
\eeq
Combining the two above constraints yields the
relativistic Poisson-like equation
\beq
\frac{1}{a^2}{\vec \nabla}^2\left[\psi+H(a^2E'-aB)\right]=4\pi G\,
\delta\epsilon, \label{epsilon_ls}
\eeq
which can also be directly obtained by linearizing the spatial
components of Eq.~(\ref{poisson2}). This equation  shows that the comoving
energy density perturbation $\delta\epsilon$ is second order in
the spatial gradients, and thus negligible on large scales in
Eq.~(\ref{s_evol_1}).

In contrast with $\sf$, the quantity $\sif$ is not gauge-invariant.
This is why it is useful to consider the gauge invariant
Sasaki-Mukhanov variable $Q_{\rm SM}$, defined as \cite{SM}
\beq
Q_{\rm SM} \equiv \sif + \frac{\sib' }{H} \psi.
\eeq
Note that the above traditional definition does not follow exactly from our definition
  of  $Q_a$ given earlier in
Eq.~(\ref{Q_two}). Indeed, from $Q_a$,  one can extract a scalar quantity $Q$ defined
as
\beq
Q_i = \partial_i Q, \qquad Q \equiv \sif - \frac{\sib' }{H} \af ,
\eeq
where $\af$ can be written in terms of metric perturbations by
making use of Eqs.~(\ref{alpha_def}) and (\ref{delta_Theta})
(see \cite{Langlois:2005qp}),
\beq
\af = - \psi + \frac{1}{3} \int \vec \nabla^2 (E'-B/a) dt.
\label{alpha_psi}
\eeq
Thus the scalar variable $Q$ coincides with $Q_{\rm SM}$ only in
the large-scale limit.

In the flat gauge, defined by $\hat \psi \equiv 0$, $\sif$
coincides with $Q_{\rm SM}$,
\beq
\hat \sif =Q_{\rm SM}.
\eeq
In this gauge, it is possible to use the momentum constraint
equations (\ref{energy_constraint_1}) to derive the metric
perturbation $A$ as a function of $Q_{\rm SM}$,
\beq
\hat A = -\frac{H'}{H \sib'}Q_{\rm SM}, \label{Af1}
\eeq
and one can write the Poisson equation (\ref{epsilon_ls}) as
\beq
\vec \nabla^2 (\hat E'-\hat B/a) =-\frac{H'}{H\sib'} \left[Q_{\rm
SM}' + \left( \frac{H'}{H} - \frac{\sigma''}{\sigma'} \right) Q_{\rm SM}
- 2 \thetab' \sf \right] , \label{Qf'}
\eeq
where we have used the expression (\ref{delta_epsilon_1})
specialized to the flat gauge. By replacing Eqs.~(\ref{Af1}) and
(\ref{epsilon_ls}) into the evolution equation of $\hat \sif$, one
finds the evolution equation of $Q_{\rm SM}$
\cite{TN,Gordon:2000hv},
% change FV 22-10-2006
\bea
&&Q_{\rm SM}''+ 3H Q_{\rm SM}' + \left[\bar V_{, \sigma \sigma} -
\thetab'{}^2 - 2\frac{H'}{H}  \left(
  \frac{\bar V_{, \sigma}}{\sib'} +
\frac{H'}{H} - \frac{\sib''}{\sib'}\right) - \frac{\vec \nabla^2}{a^2} \right] Q_{\rm SM} \nonumber \\
&& \quad = 2 (\thetab' \sf)' -2 \thetab'\left(\frac{\bar V_{,
\sigma}}{\sib'} + \frac{H'}{H}  \right) \sf. \label{equation_Q_1}
\eea
% end of change

When one considers only {\em large scales},
the expression (\ref{Qf'}) reduces to
\beq
Q_{\rm
SM}' + \left( \frac{H'}{H} - \frac{\sigma''}{\sigma'} \right) Q_{\rm SM}
- 2 \thetab' \sf \approx 0,
\label{integral_Q_1}
\eeq
which means that there exists a first integral for the quantity
$Q_{\rm SM}$ and that the second-order equation of motion
(\ref{equation_Q_1}) is not necessary in this limit. In fact, one
can easily check that the large-scale limit of
(\ref{equation_Q_1}) is an automatic consequence of the first
integral (\ref{integral_Q_1}).

Let us now consider  the evolution equation for $\zeta_a$ with two
scalar fields, Eq.~(\ref{zeta_evolution_lin}). The  spatial
components of $\zeta_a$, at linear order, are given by
\cite{Langlois:2005qp}
\beq
\delta \zeta_i =\partial_i \zeta, \qquad \zeta \equiv \af
-\frac{H}{\rhob'} \rhof. \label{zeta_first}
\eeq
From Eq.~(\ref{alpha_psi}) the scalar variable $\zeta$ is thus
related to the Bardeen gauge invariant variable $\zeta_{\rm B}$,
defined as \cite{Bardeen:1980kt,Bardeen:1983qw}
\beq
\zeta_{\rm B} \equiv -\psi - \frac{H}{\rhob'} \rhof,
\eeq
in a similar way to how $Q$ is related to $Q_{\rm SM}$, and on large
scales these two quantities coincide.

According to Eq.~(\ref{delta_dot_X}), the spatial components of
$\dot\zeta_a$ are
\beq
\delta( \dot \zeta_i) = \partial_i \zeta^{\,\prime}.
\eeq
One then finds  that the spatial components of
Eq.~(\ref{zeta_evolution_lin}) correspond to the spatial gradient
of
\beq
\zeta' = \frac{2 \bar V_{,\sigma}}{3 \sib'{}^3}  \delta \epsilon -
\frac{2H}{\sib'}\bar \theta' \delta s + \frac{1 }{3 \sib' a^2}
\vec\nabla^2 \sif. \label{zeta_evolution_lin_coordinates}
\eeq
On large scales, the spatial gradients can be neglected and the
above equation reduces to the well-known relation
\beq
\zeta' \approx -
\frac{2H}{\sib'}\bar \theta' \delta s.
\eeq
Eq.~(\ref{zeta_evolution_lin_coordinates}) can be easily specialized
to the case of a single
scalar field by simply setting $\delta \sigma=\delta \phi$ and
$\delta s=0$. One recovers in particular that on large scales
$\zeta$ is conserved.
Note also that from Eq.~(\ref{zeta_R_two}) there is
a simple relation between $\zeta$ and $\R$,
\beq
\zeta+\R=-\frac{\ab'}{\rhob'} \delta \epsilon,
\label{relation_R_zeta}
\eeq
which shows that $\zeta$ and $-\R$ coincide on large scales.

In conclusion, all the equations in this section, derived directly
from our covariant formalism, exactly reproduce the linear results
of \cite{Gordon:2000hv}. In the next section we turn to the
second-order perturbations and derive novel results.

%%%%%%%%%%%%%%%%%%%%%%%%%%%%%%%%%%%%%%%%%%%%%%%%%%%%%%%%%%%%%%%%%%%%%%%%%%%%%%%%%%%%%%%
%%%%%%%%%%%%%%%%%%%%%%%%%%%%%%%%%%%%%%%%%%%%%%%%%%%%%%%%%%%%%%%%%%%%%%%%%%%%%%%%%%%%%
\section{Second order perturbations}
\label{sec:secondorder}
%%%%%%%%%%%%%%%%%%%%%%%%%%%%%%%%%%%%%%%%%%%%%%%%%%%%%%%%%%%%%%%%%%%%%%%%%%%%%%%%%%%%%%
%%%%%%%%%%%%%%%%%%%%%%%%%%%%%%%%%%%%%%%%%%%%%%%%%%%%%%%%%%%%%%%%%%%%%%%%%%%%%%%%%%%%%%%%
\def\Xf{X_1}
\def\Xs{X_2}
\def\af{{\delta \alpha}}
\def\as{{\delta \alpha}^{(2)}}
\def\sif{{\delta \sigma}}
\def\sis{{\delta \sigma}^{(2)}}
\def\sf{{\delta s}}
\def\ss{{\delta s}^{(2)}}
\def\phf{{\delta \varphi}}
\def\phs{{\delta \varphi}^{(2)}}
\def\phif{{\delta \phi}}
\def\chif{{\delta \chi}}
\def\phis{{\delta \phi}^{(2)}}
\def\chis{{\delta \chi}^{(2)}}
\def\sif{{\delta \sigma}}
\def\phis{{\delta \phi}^{(2)}}
\def\rhof{{\delta \rho}}
\def\rhos{{\delta \rho}^{(2)}}
\def\ps{{\delta P}^{(2)}}
\def\pf{{\delta P}}
\def\ab{\bar{\alpha}}
\def\rhob{\bar{\rho}}
\def\pb{\bar{P}}
\def\sib{\bar{\sigma}}
\def\phib{\bar{\phi}}
\def\chib{\bar{\chi}}
\def\sb{\bar{s}}
\def\thetab{\bar{\theta}}
\def\As{A^{(2)}}
\def\psis{\psi^{(2)}}
\def\Qs{Q^{(2)}}
\def\Qf{Q}
\def\Vi{V_i}

In this section we consider second-order perturbations. We will
thus decompose any scalar quantity $X$ as
\beq
X(t,x^i) \equiv \bar X(t)+\delta X^{(1)}(t,x^i) + \delta
X^{(2)}(t,x^i) \label{sec_order_dec},
\eeq
where $\bar X(t)$ is the background part and $\delta X^{(1)}$ and
$\delta X^{(2)}$ are respectively the first and second-order
contributions (note that we do not follow here the convention of
including a numerical factor $1/2$ in front of the second-order
contribution). In our subsequent equations, to simplify the
notation, we will often omit the index ${}^{(1)}$ for the
first-order quantities, unless it is required for clarity reasons.

Our main purpose will be to expand our equations governing
$\sigma_a$, $s_a$ and $\zeta_a$ at second order in the
perturbations. Before undertaking this task, it is instructive to
recall the results of our previous works, in particular of
\cite{Langlois:2005qp}, on the second-order component of $\zeta_a$
for a general fluid. For a discussion on the
expansion  of $\zeta_i$ and the gauge invariance of $\zeta$ at order
higher than second see \cite{Enqvist:2006fs}.

\subsection{$\zeta_a$ at second order and the issue of gauge-invariance}

As shown in  \cite{Langlois:2005qp}, the second-order expression
of the spatial components $\zeta_i$ can be written, after some
manipulations,  in the form
\beq
\zeta^{(2)}_i=\partial_i\zeta^{(2)}+\frac{\rhof}{\rhob'}
\partial_i\zeta^{(1)}{}'
\label{delta_zeta_2}
\eeq
with
\beq
\zeta^{(2)}\equiv \as -\frac{H}{\rhob'}\rhos
 - \frac{\rhof}{\rhob'} \left[ \zeta^{(1)}{}' +
\frac{1}{2}\left(\frac{H}{\rhob'}\right)'  \rhof \right]
 ,
\eeq
and $\zeta^{(1)}$ being given in Eq.~(\ref{zeta_first}). In the
large-scale limit, i.e., keeping only the scalar perturbations
without gradients, where
\beq
\alpha\approx {\rm ln}\, a -\psi-\psi^2, \label{psi_alpha}
\eeq
we have also shown that our second-order quantity $\zeta^{(2)}$
could be easily related to the second-order quantity defined by
Malik and Wands \cite{Malik:2003mv},
which we denote here by $\zeta^{(2)}_{\rm MW}$,
namely
\beq
\zeta^{(2)}\approx\zeta^{(2)}_{\rm MW}-{\zeta^{(1)}_{\rm MW}}^2.
\eeq
Since $\zeta^{(2)}_{\rm MW}$ was constructed by explicitly
requiring {\it gauge-invariance}, this implies that, on large
scales, our $\zeta^{(2)}$ also behaves like a gauge-invariant
quantity. It is instructive to understand directly why our
$\zeta^{(2)}$ is indeed gauge-invariant on large scales.

Under a second-order  coordinate transformation,
\beq
x^\mu\rightarrow \tilde{x}^\mu=x^\mu-\xi_{(1)}^\mu+\frac{1}{2}
\xi_{(1)}^\nu \xi_{(1),\nu}^\mu- \xi_{(2)} ^\mu,
\eeq
generated  by the vector fields
$\xi_{(1)}^a$ and $\xi_{(2)}^a$, the first and second-order
perturbations of a tensor  ${\bf T}$ transform as
\cite{second_order}
\beq
{\bf \delta T}^{(1)}\rightarrow {\bf \delta
T}^{(1)}+\L_{\xi_{(1)}} {\bf T}^{(0)},\quad {\bf \delta
T}^{(2)}\rightarrow {\bf \delta T}^{(2)}+\L_{\xi_{(2)}}{\bf
T}^{(0)}+\frac{1}{2}\L_{\xi_{(1)}}^2{\bf T}^{(0)}+ \L_{\xi_{(1)}} {\bf
\delta T}^{(1)}. \label{gauge_transformation}
\eeq
Since $\zeta_a$ vanishes at zeroth order, $\zeta_a$ is
automatically gauge-invariant at {\em first order}, according to
the first expression above. However, $\zeta_a$ is not
gauge-invariant at second order and
 the corresponding gauge transformation is given, according to
Eq.~(\ref{gauge_transformation}), by
\beq
\zeta_a^{(2)}\rightarrow \zeta_a^{(2)}+ \L_{\xi_{(1)}}
\zeta_a^{(1)}.
\eeq
Concentrating now on large scales, one finds from the expression of the Lie derivative that
\beq
\L_{\xi_{(1)}}\zeta_i\approx\xi_{(1)}^0\partial_0\zeta_i^{(1)},
\eeq
 where we have neglected the terms of higher
order in spatial gradients. Consequently, the second-order spatial
components of $\zeta_a$ transform on {\em large scales} according
to
\beq
 \zeta_i^{(2)} \ \rightarrow \ \zeta_i^{(2)}+
\xi^0_{(1)}\partial_i\zeta^{(1)}{}' \qquad {\rm (large \
scales)}\, .
\eeq
By noting that, at first order,
\beq
\frac{\rhof}{\rho'} \ \rightarrow \ \frac{\rhof}{\rho'}+
\xi^0_{(1)},
\eeq
it is easy to see that the quantity
\beq
\zeta^{(2)}_i -\frac{\rhof}{\rho'}
\partial_i \zeta^{(1)}{}',
\eeq
or equivalently $\partial_i \zeta^{(2)}$,
is gauge-invariant at second order, on large scales. Thus, this
proves that $\zeta^{(2)}$ defined above is indeed gauge-invariant
on large scales.

\subsection{Adiabatic and entropy fields}

Here we derive the evolution equations for the adiabatic and
entropy field perturbations  at {\em second order}. For
simplicity, we will restrict ourselves to {\em large scales} and
we will thus start from the equations expanded in spatial
gradients discussed in Sec.~\ref{sec:ls}. For convenience, we have
collected in the appendix various background and first-order
expressions that will be used in the rest of this section.

The second-order evolution for the perturbations of a {\em single}
scalar field, in a coordinate based approach, has been considered
in
\cite{Acquaviva:2002ud,Noh:2004bc,Bartolo,Vernizzi:2004nc,Finelli:2006wk,Nakamura}.
The multi-field case, in the large-scale limit, has been studied
in detail by Malik in \cite{Malik:2005cy}  (see also
\cite{Enqvist:2004bk,Anupam,Cline}) and, using the separate
universe approach, in \cite{Lyth:2005fi,Vernizzi:2006ve}. However,
the second-order decomposition into adiabatic and entropy
components has not been given. This decomposition, which appears
quite involved  in the coordinate based approach, becomes natural
in our nonlinear formalism for scalar fields and can be derived
straightforwardly since we have already identified the fully
nonlinear adiabatic and entropy components.

We start by expanding the definition of $\sigma_a$ in
Eq.~(\ref{tan_ort1}) at second order. After some straightforward
manipulations, one can write the spatial components of $\sigma_i$
as
\beq
\delta \sigma_i^{(2)} =  \frac{1}{\sib'} \left[ \partial_i (\phib' \phis +
\chib' \chis )
+\left( \sf'
+  \bar \theta'  \sif \right)\partial_i \sf \right].  \label{1st_deltasigma}
\eeq
To deal with the term  $\sf' \partial_i \sf$, which cannot be
written as a total gradient, it is convenient to introduce the
spatial vector
\beq
V_i \equiv \frac{1}{2} (\sf \partial_i \sf' - \sf' \partial_i \sf) =
-\sf' \partial_i \sf + \frac{1}{2} \partial_i (\sf \sf'), \label{V_def}
\eeq
which vanishes when $\sf'$ and $\sf$ have the same spatial
dependence, i.e., $\sf'=f(t)\sf$.

 By expanding also
the definition of $s_a$ in  Eq.~(\ref{tan_ort2}), one finds,
for $s_i$ and $\sigma_i$, respectively,
\bea
\label{sigi2} \delta \sigma_i^{(2)} \tackl = \tackr \partial_i
\sis + \frac{\thetab'}{\sib'} \sif\partial_i \sf -\frac{1}{\sib'} V_i, \label{si_i} \\
\label{s_i}  \delta s_i^{(2)}  \tackl = \tackr \partial_i \ss +
\frac{\sif}{\sib'}  \partial_i \sf',
\eea
with
\bea
\label{si2} \sis \tackl \equiv \tackr \frac{\phib'}{\sib'} \phis +
\frac{\chib'}{\sib'} \chis + \frac{1}{2 \sib'} \sf \sf',  \\
\ss \tackl \equiv \tackr - \frac{\chib'}{\sib'} \phis +
\frac{\phib'}{\sib'} \chis -\frac{\sif}{\sib'}  \left( \sf' +
\frac{\bar \theta'}{2}
 \sif\right).\label{sis}
\eea
The form of the right hand side of Eq.~(\ref{s_i}) has been chosen
by analogy with the form (\ref{delta_zeta_2}). Since $s_a$
vanishes at zeroth order, arguments similar to those of the
previous subsection ensure that $\ss$, defined in Eq.~(\ref{sis}),
is gauge invariant on large scales. The form of $\ss$ is in some
sense dictated by our  covariant definition.

Note that  $\delta s_i^{(2)}$ contains the first-order adiabatic
perturbation. This is due to the fact that the adiabatic and
entropy components are defined locally: whereas the first-order
components are defined with respect to a background basis in field
space, which is {\em only} time dependent, the second-order
components will be sensitive to the first-order fluctuations of
the field space basis, which can be expressed in terms of the
first-order adiabatic and entropy components. The adiabatic
component $\sigma_a$ does not vanish at zeroth order and $\sis$ is
not a gauge invariant variable. Our formalism does not dictate its
form, which is thus chosen at our convenience.

Let us now  discuss  further the adiabatic component. Since $\sis$
is not gauge-invariant on large scales, in contrast with $\ss$, it
is useful to consider our generalization of the Sasaki-Mukhanov
variable, $Q_a$,  defined in Eq.~(\ref{Q_two}). Its spatial
components can be expanded at second order in the perturbations,
similarly to what we have done with $\zeta_a$,
\beq
Q_i^{(2)} = \partial_i \Qs + \frac{\af}{H}  \partial_i \Qf'
+  \frac{\thetab'}{\sib'} \Qf\partial_i \sf -\frac{1}{\sib'}V_i, \label{Q2_i}
\eeq
where $\Qs$ is defined as
\beq
\Qs \equiv \sis - {\sib'\over H}\as
 - \frac{\af}{H} \left[ Q^{(1)}{}' + \frac{1}{2}
 {\left({\sib'\over H}\right)}' \af - \thetab' \sf \right]
\label{Q_2}.
\eeq
From this expression and Eq.~(\ref{psi_alpha}) it is natural to define
\beq
Q_{\rm SM}^{(2)} \equiv \sis + {\sib'\over H} (\psis+ \psi^2)
 + \frac{\psi}{H} \left[ Q_{\rm SM}^{(1)}{}' - \frac{1}{2}
 {\left({\sib'\over H}\right)}'
 \psi -  \thetab' \sf \right]
\label{QSM_2}
\eeq
as the  local part of the scalar gauge invariant second-order
Sasaki-Mukhanov variable.  Restricted to a single scalar field,
this definition coincides with the one given in
\cite{Malik:2005cy}.
 Note that we cannot write Eq.~(\ref{Q2_i})
in the same form as Eq.~(\ref{delta_zeta_2}) because the last two
terms on the right hand side cannot be written as a total spatial
gradient.

The second order (in time) evolution of $\sigma_a$ is given by
Eq.~(\ref{sigma2}). However, on large scales we do not need to
compute a second-order differential equation because the
adiabatic evolution is governed by a first integral, as in the
linear case. This first integral is obtained directly from the
constraint equations and it is not necessary to expand
(\ref{sigma2}) at second order in the perturbations.

In order to compute this first integral we need  the
second-order energy and momentum constraints, which can be derived
by expanding Eqs.~(\ref{e_c}) and (\ref{m_c}) and by using
\beq
\delta \Theta^{(2)} \approx  \frac{9}{2} H A^2 +3A \psi' -6 \psi
\psi'.
\eeq
On large scales, one can write the second-order energy constraint
equation as
\beq
3 H\left[H \As +  \psi^{(2)}{}' +2  \psi \psi'
- \frac{1}{2} \frac{\psi'{}^2}{H}
-2 A(HA+ \psi' )   \right] \approx -4
\pi G \rhos,
\label{energy_2}
\eeq
where $\rhos$ is given by
\bea
\rhos \tackl \approx \tackr \sib' \sis{}' - \sib'^2 \As
+ \bar V_{,\sigma} \sis + 2 \bar V_{,s}\ss \nonumber \\ \tackl - \tackr
2 A ( \sib' \sif'
 - \sib'{}^2 A - \sib' \thetab' \sf) + \Delta_\rho,
\eea
and $\Delta_\rho$ is a quadratic function of $\delta
\sigma^{(1)}$, $\delta s^{(1)}$, and their first derivatives,
given by
\bea
\Delta_\rho\tackl  \equiv  \tackr \frac{1}{2} \sif' (\sif'-2
\thetab' \sf) -  \sf' \left(\frac{V_{,\sigma}}{\sib'}
\sf +  \thetab'
\sif\right)
\nonumber \\
\tackl + \tackr  \frac{1}{2} (\bar V_{,\sigma \sigma} -
\thetab'{}^2 ) \sif^2 + (\bar V_{, ss} + 2 \thetab'{}^2 )
\sf^2 + \bar V_{,s\sigma} \sf \sif.
\eea

The second-order momentum constraint equation reads
\beq
\partial_i \left[H \As + \psi^{(2)}{}' + 2 \psi \psi' - \frac{1}{2}HA^2
-A(HA+\psi') \right]
\approx -4 \pi
G \delta q^{(2)}_i,
\label{momentum_2}
\eeq
where the second-order momentum $\delta q^{(2)}_i$ is given, from
Eq.~(\ref{qa}), by
\beq
\delta q_i^{(2)} =-\partial_i \left(\sib' \sis
+\frac{1}{2} \frac{\sib''}{\sib'} \sif^2 + \thetab' \sif \sf
\right) -\frac{1}{\sib'}
\delta \epsilon \partial_i \sif +V_i.
\label{momentum_2_def}
\eeq
As already noticed in \cite{Rigopoulos:2005xx} for large scales,
 $\delta q_i^{(2)}$ {\em cannot} be written as a total gradient when several
scalar fields are present. After neglecting $\delta \epsilon$ on
large scales in  the above equation, this is manifest because of
the presence of $V_i$. This implies that,  in principle, if $V_i$
does not vanish,  one cannot define at second-order a comoving
gauge,  i.e., such that $\delta q_i^{(1)}=0$ {\em and} $\delta
q_i^{(2)}=0$, in  contrast with the linear theory or the
single-field case.

However, it is instructive to  derive the  evolution equation for $V_i$ on
large scales by using the linear evolution equation for $\sf$,
Eq.~(\ref{s_evol_1}), neglecting the gradient term and $\delta
\epsilon$ at first order. One finds
\beq
V_i'+ 3 H V_i=0,
\label{evolution_Vi}
\eeq
which implies that, in an expanding universe, $V_i$ will  decay
like $a^{-3}$ and rapidly become negligible even if it is nonzero
initially.  Consequently, in
an expanding universe one can in practice  ignore $V_i$ on large
scales and thus define,  {\em in an approximate sense}, a comoving
gauge at second order, which coincides with $\sif^{(1)}=0=\sis$.
In this approximate comoving gauge, the momentum
(\ref{momentum_2_def}) can be written as a total gradient. In the
rest of the paper, in order to remain as general as possible, we
will keep the term $V_i$.\footnote{Note that a similar term
(defined for several scalar fields) has appeared in a recent paper  
\cite{Seery:2006vu} where it
was neglected. Here we have provided the explanation for why it can be
neglected.}

Similarly to the first-order case, it is possible to combine the
energy and momentum constraint equations and derive the
relativistic Poisson-like equation analogous to
Eq.~(\ref{epsilon_ls}), which corresponds to the expansion, at
second order and on large scales, of Eq.~(\ref{poisson2}). By
expanding Eq.~(\ref{epsilon_ls_tilde}) and using
(\ref{eps_eq_eps_tilde}), one has
\beq
\delta \epsilon^{(2)}_i \approx \partial_i \rhos - 3H \delta q_i^{(2)}
- \delta \Theta \delta q_i \approx 0, \label{Poisson_2}
\eeq
where the last approximate equality is a consequence of
Eqs.~(\ref{delta_Theta}) and (\ref{energy_2})  and confirms our
conclusion of  Sec.~\ref{sec:ls} in a covariant context, namely
that we can neglect $\epsilon_a$ on large scales. The second-order
spatial components of  $\epsilon_a$ defined in
Eq.~(\ref{epsilon}), can be decomposed as
\beq
\delta \epsilon_i^{(2)} = \partial_i \delta \epsilon^{(2)} +
\frac{\sif}{\sib'}
\partial_i \delta \epsilon^{(1)}{}' -3H V_i, \label{epsilon_i2}
\eeq
with $\delta \epsilon^{(2)}$ defined by
\beq
\delta \epsilon^{(2)} \equiv \rhos - {\rhob'\over \sib'} \sis
 - \frac{\sif}{\sib'} \left[ \delta \epsilon^{(1)}{}'  +\frac{1}{2}
 {\left({\rhob'\over \sib'}\right)}'
 \sif  +    \frac{\rhob'}{\sib'} \bar \theta' \sf \right]
 . \label{delta_epsilon_def_2}
\eeq
It is only when $V_i$ is negligible that  the quantity $\delta
\epsilon^{(2)}$ can be interpreted as  the comoving energy density
at second order. Otherwise, as discussed before, the comoving
gauge cannot be defined.

Using  the decomposition (\ref{epsilon_i2}) and the fact that
$\delta \epsilon^{(1)}$ is negligible on large scales,
Eq.~(\ref{Poisson_2}) can be written as
\beq
\partial^{2} \delta \epsilon^{(2)} \approx 3H
\partial^iV_i.
\label{epsilon2_ls}
\eeq
When $V_i$ is negligible, and only then, one finds that, like at
first order, the second-order comoving energy density is
negligible on large scales,
\beq
\delta \epsilon^{(2)} \approx 0.
\eeq

Having discussed the general properties of the second-order
constraint equations, we can now derive the evolution equation of
the gauge-invariant adiabatic component $Q_{\rm SM}^{(2)}$.
Similarly to what we have shown in the previous section for the
first-order variables, the simplest way to derive  an equation
satisfied by $Q_{\rm SM}^{(2)}$ is to work in the flat gauge $\hat
\psi^{(1)}= 0 =\hat \psi^{(2)}$, where $\sis$ reduces to $Q_{\rm
SM}^{(2)}$,
\beq
\label{Q_3} \hat \sis = Q_{\rm SM}^{(2)}.
\eeq
In this gauge, Eqs.~(\ref{energy_2}) and (\ref{momentum_2})
reduce, respectively, to
\bea
3 H^2\left(\hat  \As  -2 \hat  A^2  \right) \tackl \approx \tackr-4
\pi G \hat \rhos, \\
H \partial_i \left( \hat  \As  - \frac{3}{2} \hat  A^2 \right)
\tackl \approx \tackr -4 \pi
G \hat{\delta q}^{(2)}_i.
\eea
Using the first-order constraint equations Eqs.~(\ref{Af1}) and
(\ref{integral_Q_1}), these can be rewritten as
\bea
\hat \As \tackl \approx \tackr -
\frac{\sib'}{2\bar V} \left[ Q^{(2)}_{\rm SM}{}'
+ \frac{\bar V_{,\sigma}}{\sib'} \left( Q^{(2)}_{\rm SM} - 2
\frac{H'}{H \sib'} Q_{\rm SM}^2\right) \right. \nonumber \\ \tackl - \tackr
\left. 2
\thetab' \left(  \ss -
\frac{H'}{H \sib'}
Q_{\rm SM} \sf \right)
+\frac{\hat \Delta_\rho}{\sib'} \right],
\label{c_1}
\eea
and
\beq
\hat \As \approx -\frac{H'}{H \sib'}  \left[ \Qs_{\rm SM} +
\frac{\thetab'}{\sib'} Q_{\rm SM} \sf + \frac{1}{2 \sib'} \left(
\frac{\sib''}{\sib'} -3 \frac{H'}{H} \right) Q_{\rm SM}^2 -
\frac{1}{\sib'} \partial^{-2}
\partial^i V_i\right]  . \label{c_2}
\eeq
 This last equation
contains a nonlocal term because we have written the momentum
constraint as a scalar equation while keeping the second-order
vector $V_i$ defined in (\ref{V_def}). By combining these two
relations to get rid of $\As$ one obtains the following first
integral for $Q^{(2)}_{\rm SM}{}$:
\bea
Q_{\rm SM}^{(2)}{}' \tackl + \tackr
\left(\frac{H'}{H}-\frac{\sib''}{\sib'} \right) Q_{\rm SM}^{(2)}
-2 \thetab' \ss \nonumber \\ \tackl \approx \tackr  - \frac{1}{\sib'} \left[
-2 \frac{H'}{H} \frac{\bar V_{,\sigma}}{\sib'} + \frac{1}{2}
\left(\frac{\sib''}{\sib'} -3 \frac{H'}{H} \right)\left(3 H +\frac{H'}{H} \right) \right] Q_{\rm SM}^2
- \frac{\hat \Delta_\rho}{\sib'}
\nonumber \\  \tackl - \tackr 3 \frac{\thetab'}{\sib'}
\left(H+\frac{H'}{H} \right) Q_{\rm SM} \sf +
\frac{1}{\sib'}
\left(3 H+\frac{H'}{H} \right) \partial^{-2} \partial^i V_i.
\label{integral_Q_2}
\eea
This equation is the second-order equivalent of
(\ref{integral_Q_1}).  We have put together on the left hand side
all the terms depending on purely second-order quantities, and on
the right hand side all the terms which are quadratic in
first-order quantities. As in the first-order case, the entropy
perturbation sources the evolution of the adiabatic perturbation.
The nonlocal term containing $V_i$ comes from the momentum
constraint and is a new feature with respect to the first-order
case (or the second-order case for a {\em single} scalar field).
However, as we have discussed earlier, it becomes quickly
negligible on large scales in an expanding universe, in which case
the first integral (\ref{integral_Q_2}) becomes a scalar local
equation.

It is also possible to construct a second-order (in time)
differential equation for $Q^{(2)}_{\rm SM}{}$ that {\em always} looks
 purely local,  even in the presence of a
non-vanishing $V_i$. Using the evolution equation of $V_i$,
Eq.~(\ref{evolution_Vi}), this can be done by taking an
appropriate linear combination  of (\ref{c_2}) and its time
derivative so that all nonlocal terms  cancel each other. One then
obtains a scalar equation for a linear combination of $A^{(2)}{}'$
and $\As$ which can be transformed into a second-order
differential equation for $Q^{(2)}_{\rm SM}{}$ by substituting the
expression for $\As$ given by the energy constraint.\footnote{Note
that a similar second-order equation was obtained in
\cite{Malik:2005cy} by using the second-order constraint from the
diagonal space-space components of Einstein's equations. It can be
shown however that this constraint is redundant with the energy
and momentum constraints which we are using here and the two
procedures are thus essentially equivalent.} But, even then, the
initial conditions for this second-order differential equation
must be compatible with Einstein's equations and thus satisfy the
nonlocal constraint (\ref{integral_Q_2}). As a consistency check,
we have also verified that this second-order (in time) equation is
equivalent to the equation obtained by expanding to second-order
(in the perturbations) the equation (\ref{sigma2}) for $\sigma_a$,
when combined with the constraints (\ref{c_1}--\ref{c_2}).

 We can now derive the evolution equation for the entropy perturbation
$\ss$, which will be second order in time. Since we are
restricting ourselves to {\em large scales}, we simply expand the
spatial components of Eq.~(\ref{s4}) up to second order. This
gives
%FVerrata
\bea
&&\delta (\ddot s_i)^{(2)}+3H \delta (\dot s_i)^{(2)}+ \left(\bar
V_{,ss}+3{\bar {\theta}}^{\prime 2}\right)\delta s_i^{(2)} +\delta
\Theta
\partial_i \sf'\cr && \quad + \left[\delta
V_{ss}+6\bar{\theta}'\delta(\dot\theta)\right]\partial_i \sf
\approx - 2 \frac{\bar \theta'}{\sib'} \delta \epsilon^{(2)}_i ,
\label{pre_s_evol_2}
\eea
where we have neglected the gradient of the comoving energy
density at first order, $\delta \epsilon_i^{(1)}$ which according
to Eq.~(\ref{epsilon_ls}) is subdominant on large scales.

To proceed, we need the spatial components of the first and second
time derivatives of the covectors $s_a$. By using Eq.~(\ref{Lie2})
for $s_i$ at second order, and ignoring the higher-order terms in
the gradient expansion,  one obtains
\beq
\label{dotsi2} \delta (\dot s_i)^{(2)} \approx \delta
s_i^{(2)}{}'- A
\partial_i \sf',
\eeq
and, by applying once more (\ref{Lie2}),
\beq
\label{ddotsi2} \delta (\ddot s_i)^{(2)} \approx  \delta
s_i^{(2)}{}'' - A'
\partial_i \sf' -2 A
\partial_i \sf'' .
\eeq
We can then substitute the expressions (\ref{s_i}) and
 (\ref{dotsi2}--\ref{ddotsi2}) for the second-order entropy
component and its derivatives. Using  various first-order
expressions  summarized in the appendix, we finally find that the
equation (\ref{pre_s_evol_2}) can be written as the spatial
gradient of the following scalar equation
\bea
&&\ss{}''+3H \ss{}'+\left(\bar V_{,ss}+3{\bar {\theta}}^{\prime
2}\right)  \ss \approx   -\frac{\bar \theta'}{\sib'}  \sf'{}^2
\nonumber \\
&& \qquad - \frac{2}{\sib'}\left( \bar \theta''+ \bar \theta'
\frac{\bar V_{,\sigma}}{\sib'} -  \frac{3}{2} H \thetab'\right)
\sf \sf'
\nonumber \\
&& \qquad - \left( \frac{1}{2} \bar V_{,sss} - 5\frac{\bar
\theta'}{\sib'} \bar V_{,ss} - 9 \frac{\bar \theta'{}^3}{\sib'}
\right)\sf^2 - 2  \frac{\thetab'}{\sib'} \delta \epsilon^{(2)} .
\label{s_evol_2}
\eea
Note that this equation is closed, in the sense that only the
entropy field perturbation appears: even when $\delta
\epsilon^{(2)}$ is not negligible, it  can be written in terms 
of $\sf$ and $\sf'$ by using Eqs.~(\ref{epsilon2_ls}) and
(\ref{V_def}). Thus, on large scales the entropy field evolves
independently of the adiabatic components, as in the linear
theory.

\subsection{Generalized uniform density and comoving curvature perturbations}

We now derive the large-scale evolution equation for
$\zeta^{(2)}$, expanding at second order Eq.~(\ref{zeta_ls}). In
the fluid description, it was shown in
\cite{Malik:2003mv,Langlois:2005qp} that on large scales the
evolution equation for $\zeta^{(2)}$ can be written as
\beq
\zeta^{(2)}{}' \approx - \frac{H}{\bar \rho+ \bar P} \Gamma^{(2)}
- \frac{1}{\bar \rho+ \bar P} \Gamma_1 \zeta_1'   ,
\label{conserv_zeta2}
\eeq
where $\Gamma^{(2)}$ can be read from the second-order
decomposition of the quantity $\Gamma_a$ for a fluid, defined in
Eq.~(\ref{Gamma_def}), i.e.,
\beq
\delta \Gamma_i^{(2)} =
\partial_i  \Gamma^{(2)} + \frac{\rhof}{\rhob'}
\partial_i \Gamma^{(1)}{}'. \label{so_form}
\eeq
In the two-scalar field case considered here we must compare this
expansion with the expression for $\Gamma_a$ given in the
large-scale limit by Eq.~(\ref{Gamma_ls}). Expanding this equation
at second order, using Eqs.~(\ref{Poisson_2}) and (\ref{s_i}), one
obtains
\beq
\delta \Gamma_i^{(2)} \approx  2\frac{\sib'}{\rhob'}\bar V_{,\sigma}
\delta \epsilon_i^{(2)}
- 2 \bar V_{,s} \left(\partial_i
\ss +\frac{\sif}{\sib'} \partial_i \sf' \right) -2 \delta V_{,s}
\partial_i \sf, \label{ss1}
\eeq
 where we have used that $\delta \epsilon^{(1)}$ is negligible
on large scales. Replacing $\delta V_{,s}$ in the last term of
Eq.~(\ref{ss1}) by the expression given in the appendix,
 one can rewrite this equation in the form
(\ref{so_form}), with
\beq
\Gamma^{(2)} \approx  2\frac{\sib'}{\rhob'}\bar V_{,\sigma}
 \delta \epsilon^{(2)}
  - 2\bar
V_{,s} \ss -  \bar V_{,ss} \sf^2 
%FVerrata
+ \frac{\bar V_{,\sigma}}{\bar \sigma'} \delta s \delta s'
.
\eeq
One can now rewrite the second-order evolution equation for
$\zeta^{(2)}$ on large scales,
 Eq.~(\ref{conserv_zeta2}), as
\beq
\zeta^{(2)}{}'\approx -\frac{H}{\sib'^2} \left[ 2\bar \theta'
\sib' \ss
 -  \left( \bar V_{,ss} + 4 \bar \theta'{}^2
\right)  \sf^2 
%FVerrata
+ \frac{\bar V_{,\sigma}}{\bar \sigma'} \delta s \delta s'
- \frac{ 2\bar V_{,\sigma}}{3H \sib'}\delta
\epsilon^{(2)}
 \right]. \label{zeta2_evol}
\eeq
 The last term on the right
hand side can be re-expressed in terms of the now familiar nonlocal
term  involving $V_i$, which decays quickly in an expanding
universe.

It is also useful to express our results in terms of $\R_a$. The
spatial components of $\R_a$ can be decomposed as
\beq
\R_i^{(2)} = \partial_i \R^{(2)} + \frac{ \sif}{\sib'}
\partial_i \R^{(1)}{}' -  \frac{H}{\sib'{}^2}V_i , \label{R_2_i}
\eeq
with
\beq
\R^{(2)} \equiv-  \as + {H\over \sib'}\sis
 + \frac{\sif}{\sib'} \left[ -\R^{(1)}{}' + \frac{1}{2}
 {\left({H\over \sib'}\right)}'
 \sif  +  \bar \theta' \frac{H}{\sib'}
 \sf \right]
\label{R_2}.
\eeq
The last term in Eq.~(\ref{R_2_i}) comes from the fact that, like
$\epsilon_a$ and in contrast to $\zeta_a$, $\R_a$ is defined in
terms of the spatial momentum which cannot be expressed in general
as a pure gradient. When this term can be neglected, and only
then, $\R^{(2)}$ coincides with the second-order comoving
curvature perturbation defined in
\cite{Maldacena:2002vr,Vernizzi:2004nc}.

It is easy to derive a first-order (in time) evolution equation
for $\R^{(2)}$ by noting that $\zeta^{(2)}$ and $\R^{(2)}$ are
related on large scales. Indeed, expanding Eq.~(\ref{zeta_R_ls})
at second order using (\ref{delta_zeta_2}) and (\ref{R_2_i}), and
neglecting terms proportional to $\delta \epsilon^{(1)}$, one gets
\beq
\zeta^{(2)} +\R^{(2)} \approx  \frac{1}{3\sib'{}^2} \delta
\epsilon^{(2)}. \label{R_zeta_rel2}
\eeq
When $\delta \epsilon^{(2)}$ is negligible on large scales, like
in an expanding universe where we can neglect $V_i$, $\zeta^{(2)}$
and $\R^{(2)}$ coincide on large scales as in the single-field
case \cite{Vernizzi:2004nc}. However, this is not true in general
in the multi-field case if $V_i$ cannot be neglected.

From this relation and the evolution equation of $\zeta^{(2)}$,
Eq.~(\ref{zeta2_evol}), one can find a large-scale evolution
equation for $\R^{(2)}$,
\beq
\R^{(2)}{}'\approx \frac{H}{\sib'^2} \left[ 2\bar \theta' \sib'
\ss
 -   \left( \bar V_{,ss} + 4 \bar \theta'{}^2
\right)  \sf^2  
%FVerrata
+ \frac{\bar V_{,\sigma}}{\bar \sigma'} \delta s \delta s'
+\left( 1 + \frac{H'}{3H^2} \right) \delta
\epsilon^{(2)} \right]. \label{R2_evol}
\eeq
 The second-order uniform adiabatic
field perturbation $\R^{(2)}$ can be related on large scales to
$Q_{\rm SM}^{(2)}$, by combining Eqs.~(\ref{QSM_2}), (\ref{R_2}),
and (\ref{psi_alpha}). One obtains
\beq
\R^{(2)} \approx \frac{H}{\sib' } \left[Q_{\rm SM}^{(2)}
-\frac{1}{\sib'} \left(Q_{\rm SM}'- \thetab'\sf \right) Q_{\rm SM}
- \frac{1}{2H} \left(\frac{H}{\sib'} \right)' Q_{\rm SM}^2
\right], \label{RQ_rel}
\eeq
 which can be
used, together with the linear first integral
(\ref{integral_Q_1}), to show that Eq.~(\ref{R2_evol}) is
equivalent to the first integral (\ref{integral_Q_2}).

\subsection{Discussion}

In the two-field system at second order, we have found several
{\em qualitatively} new  features that are absent in the linear
case or in the second-order single-field case, and that can be
viewed as purely nonlinear effects. Before concluding, it is
worth discussing these features along with the main results of this
section.

Here we have derived coupled equations governing the evolution of
the adiabatic and entropy perturbations at second order, valid in
the large-scale limit. The entropy perturbation is described by
the second-order field perturbation $\ss$, which is gauge
invariant. The adiabatic perturbation, as usual, can be described
by one out of three gauge invariant variables. Two choices are the
Sasaki-Mukhanov variable $Q_{\rm SM}^{(2)}$, defined in
(\ref{QSM_2}), and  $\R^{(2)}$, defined in Eq.~(\ref{R_2}), the
latter being related to the former via the nonlinear equation
(\ref{RQ_rel}). Another choice is the curvature perturbation on
uniform energy density hypersurfaces, represented, on large
scales, by $\zeta^{(2)}$.

The scalar evolution equations for these three va\-ria\-bles,
res\-pec\-ti\-vely Eqs.\ (\ref{integral_Q_2}), (\ref{R2_evol}) and
(\ref{zeta2_evol}), carry the same information and share the same
characteristics. They are first order (in time) and they are
sourced by the first and second-order entropy field. They also
contain a nonlocal term which appears due to the impossibility of
writing the two-field momentum as a total gradient. However, we
have shown that this nonlocal term decays rapidly in an expanding
universe. When it can be completely neglected, and only then,
$\R^{(2)}$ corresponds, on large scales, to the comoving curvature
perturbation. In this case $\R^{(2)}$ and $\zeta^{(2)}$ coincide
on large scales.

The time variation of $\R^{(2)}$ and $\zeta^{(2)}$ depends only on
the first and second-order entropy fields. Thus, the full
large-scale evolution of second-order perturbations in the
two-field system is solved by considering  the evolution equation
of either $\R^{(2)}$ or $\zeta^{(2)}$, as well as  the evolution
equations for the first order and second order entropy
perturbations, respectively Eqs.~(\ref{s_evol_1}) and
(\ref{s_evol_2}).

\section{Conclusions}
\label{sec:conclusion}

In the present work, we have developed a covariant formalism that
deals with fully nonlinear perturbations in a universe dominated
by scalar fields. In order to do so, we have introduced a unit
vector field $u^a$, which defines our time direction. In contrast
with the case of a perfect fluid where it is natural to define
$u^a$ as the fluid four-velocity, we have here left $u^a$
arbitrary. In the case of a single scalar field, it might be
convenient to take $u^a$ as the unit vector field orthogonal to
the constant scalar field hypersurfaces, so that the total
momentum vanishes, $q_a=0$. However, in the case of several scalar
fields there is in general no such choice.

For an arbitrary number of scalar fields, we have shown that it is
possible to rewrite the fully nonlinear Klein-Gordon equation in a
form which mimics  its homogeneous version. By introducing
the gradient of the scalar field, it is easy to obtain an equation
which is exact and fully nonlinear but at the same time closely
mimics the linearized Klein-Gordon equation.

We have paid special attention to the case of two scalar fields,
by introducing two combinations of the scalar field gradients
which we call adiabatic and isocurvature covectors because they
generalize the definitions introduced in the linear theory context.
Remarkably, it has been possible to derive for these covectors two
{\it exact} and {\it fully nonlinear} second-order differential
equations, where the time derivative is defined covariantly as the
Lie derivative with respect to $u^a$.  We have also derived the
evolution equation of the covariant variable $\zeta_a$ which
generalizes, in the spirit of our formalism, the curvature
perturbation on uniform energy density hypersurfaces, showing that
on large scales it is sourced only by the entropy covector.

We have used these nonlinear equations as a starting point to show
that results previously obtained in the literature with other
approaches can be derived here in a simple and straightforward
way. We have also been able to go  beyond previous works by
computing explicitly the evolution of the {\em second-order}
adiabatic and entropy components on {\em large scales}. In
particular, using the second order energy and momentum constraints
from Einstein equations, we have derived a first integral of
motion satisfied by the second-order adiabatic component, which
has the particularity to contain a nonlocal term that depends on
the first-order entropy perturbation and its time derivative.
However, this nonlocal term goes rapidly to zero in an expanding
universe  and the second order adiabatic component is, in this
limit, governed by a local first-order (in time) evolution
equation, sourced by terms depending on the second order entropy
perturbation as well as, quadratically, on the first-order entropy
perturbation. Both first and second order entropy perturbations
satisfy a second order (in time) evolution equation and the full
system of equations, valid  on large scales, is thus closed.

\appendix
\section{Useful identities in a two-field system}

\subsection{Background identities}
\beq
H'=-4 \pi G \sib^{\prime 2}.
\eeq

% change FV 22-10-2006
\bea
\bar V_{,\sigma}' \tackl = \tackr \sib' \bar V_{,\sigma\sigma}
+\thetab' \bar V_{,s},
\\
\bar V_{,s}' \tackl = \tackr \sib' \bar V_{,s\sigma}  -\bar
\theta'
\bar V_{,\sigma},\\
\bar V_{,\sigma s}' \tackl = \tackr \sib' \bar V_{,\sigma \sigma
s} + \thetab' (\bar V_{,ss} - \bar V_{,\sigma \sigma}), \\
\bar V_{,ss}'\tackl = \tackr \bar \sigma' \bar V_{,ss\sigma}  -2
\bar \theta' \bar V_{,\sigma s} .
\eea

\bea
\thetab'\tackl = \tackr-\frac{\bar V_{,s}}{\sib'}, \\
\bar \theta''\tackl = \tackr -\bar V_{,s \sigma}+\bar \theta'
\left(\frac{\bar V_{,\sigma}}{\sib'} - \frac{\sib''}{\sib'}
\right), \label{thetadotbar}\\
\bar \theta'''\tackl = \tackr - \sib' \bar V_{,\sigma \sigma s} +
\thetab'' \left(\frac{\bar V_{,\sigma}}{\sib'}-
\frac{\sib''}{\sib'} \right) + \thetab' \left(3 \bar V_{,\sigma
\sigma} -\bar V_{,ss} -2 \thetab'{}^2 - 2
\frac{\sib''}{\sib'}\frac{\bar V_{,\sigma}}{\sib'} + 3H' \right).
\label{rel_thetathree}
\nonumber \\
\eea
% end of change

\subsection{First order identities}

\bea
\delta V_{,\sigma} \tackl = \tackr \bar V_{,\sigma \sigma} \sif +
\bar V_{,\sigma s} \sf -\thetab' (\sf' +\thetab' \sif), \\
\delta V_{,s} \tackl = \tackr \bar V_{,s\sigma} \sif + \bar
V_{,ss} \sf -\frac{\bar V_{,\sigma}}{\sib'} (\sf' +\bar \theta'
\sif), \\
\delta V_{ss} \tackl = \tackr \bar V_{,ss\sigma} \sif + \bar
V_{,sss} \sf - 2 \frac{\bar V_{,s\sigma}}{\sib'} \left(\sf' +\bar
\theta' \sif \right), \label{delta_Vss} \\
\delta V_{,\sigma s} \tackl = \tackr \bar V_{,\sigma \sigma s}
\sif + \bar V_{,\sigma ss} \sf + \left( \frac{\bar V_{,s
s}}{\sib'} - \frac{\bar V_{,\sigma \sigma}}{\sib'} \right) (\sf'
+\thetab' \sif).
\eea

\bea
\delta(\dot\sigma) \tackl = \tackr \sif' - \bar \theta' \sf
-\sib'A, \\
\delta (\dot \theta) \tackl = \tackr - \frac{1}{\sib'} \left[\bar
V_{,s} A +\delta V_{,s} +\bar{\theta}'
\left( \sif'-\bar \theta' \sf \right) \right],
\\
\delta (\ddot \theta) \tackl = \tackr  -\delta V_{,\sigma s} +
\thetab' \delta \Theta +\left( \frac{\bar V_{,\sigma}}{\sib'}
-\frac{\sib''}{\sib'} \right) \delta (\dot \theta) + 2\thetab'
\frac{\delta V_{,\sigma}}{\sib'} - 2\bar \theta' \frac{\bar
V_{,\sigma}}{\sib'} \frac{\delta (\dot \sigma)}{\sib'}.
%\nonumber \\
\eea

\vspace{0.3cm}
{\bf Acknowledgments:} We acknowledge Asko Jokinen, David Lyth,
Anupam Mazumdar, Sami Nurmi, Gerasimos Rigopoulos, Bartjan van
Tent and David Wands for interesting discussions. FV wishes to
thank Ruth Durrer and the department of Theoretical Physics of the
University of Geneva, David Lyth and the department of Physics at
Lancaster during the workshop {\em Non-Gaussianity from
Inflation}, 5 -- 9 June 2006, and the Institut d'Astrophysique de
Paris for their kind hospitality while completing this work.
%FVerrata
Finally, we would like 
to thank S\'ebastien Renaux-Petel and Gianmassimo Tasinato for pointing
out a typo in Eq.~(213) and a term missing in Eqs.~(220), (221) and (225).


\begin{thebibliography}{99}

\bibitem{Langlois:2005ii}
  D.~Langlois and F.~Vernizzi,
  %``Evolution of nonlinear cosmological perturbations,''
  Phys.\ Rev.\ Lett.\  {\bf 95}, 091303 (2005)
  [arXiv:astro-ph/0503416].
  %%CITATION = ASTRO-PH 0503416;%%
  %%Cited 12 times in SPIRES-HEP

\bibitem{Langlois:2005qp}
  D.~Langlois and F.~Vernizzi,
  %``Conserved nonlinear quantities in cosmology,''
  Phys.\ Rev.\ D {\bf 72}, 103501 (2005)
  [arXiv:astro-ph/0509078].
  %%CITATION = ASTRO-PH 0509078;%%
  %%Cited 5 times in SPIRES-HEP

\bibitem{Langlois:2006iq}
  D.~Langlois and F.~Vernizzi,
  %``Nonlinear perturbations for dissipative and interacting relativistic
  %fluids,''
  JCAP {\bf 0602}, 014 (2006)
  [arXiv:astro-ph/0601271].
  %%CITATION = ASTRO-PH 0601271;%%

%%%%%%%%%%%%%%%%%%%%%%%%%%%%%%%%%%%%%%%%%%%%%%%%%%%%%%%%%%%%%%%%%%%%%%%%%%%%%
%second order
% \cite{second_order,Acquaviva:2002ud,Bartolo,Rigopoulos:2002mc,Malik:2003mv,Noh:2004bc,Vernizzi:2004nc}

\bibitem{second_order}
M.~Bruni, S.~Matarrese, S.~Mollerach and S.~Sonego,
%``Perturbations of spacetime: Gauge transformations
%and gauge invariance  at second order and beyond,''
Class.\ Quant.\ Grav.\  {\bf 14} 2585 (1997) 
[arXiv:gr-qc/9609040]; S.~Matarrese, S.~Mollerach and M.~Bruni,
%``Second-order perturbations of the Einstein-de Sitter universe,''
Phys.\ Rev.\ D {\bf 58}, 043504 (1998) [arXiv:astro-ph/9707278];
%%CITATION = ASTRO-PH 9707278;%%
M.~Bruni, F.~C.~Mena and R.~K.~Tavakol,
%``Cosmic no-hair: Non-linear asymptotic stability of de Sitter universe,''
Class.\ Quant.\ Grav.\  {\bf 19}, L23 (2002)
[arXiv:gr-qc/0107069].
%%CITATION = GR-QC 0107069;%%


%\cite{Acquaviva:2002ud}
\bibitem{Acquaviva:2002ud}
  V.~Acquaviva, N.~Bartolo, S.~Matarrese and A.~Riotto,
  %``Second-order cosmological perturbations from inflation,''
  Nucl.\ Phys.\ B {\bf 667}, 119 (2003)
  [arXiv:astro-ph/0209156].
  %%CITATION = ASTRO-PH 0209156;%%

%\cite{Maldacena:2002vr}
\bibitem{Maldacena:2002vr}
  J.~M.~Maldacena,
  % ``Non-Gaussian features of primordial fluctuations in single field
  %inflationary models,''
  JHEP {\bf 0305}, 013 (2003)
  [arXiv:astro-ph/0210603].
  %%CITATION = ASTRO-PH 0210603;%%

\bibitem{Bartolo}
  N.~Bartolo, E.~Komatsu, S.~Matarrese and A.~Riotto,
  %``Non-Gaussianity from inflation: Theory and observations,''
  Phys.\ Rept.\  {\bf 402}, 103 (2004)
  [arXiv:astro-ph/0406398].
  %%CITATION = ASTRO-PH 0406398;%%

%\cite{Rigopoulos:2002mc}
\bibitem{Rigopoulos:2002mc}
  G.~Rigopoulos,
   %``On second order gauge invariant perturbations in multi-field inflationary
  %models,''
  Class.\ Quant.\ Grav.\  {\bf 21}, 1737 (2004)
  [arXiv:astro-ph/0212141].
  %%CITATION = ASTRO-PH 0212141;%%


%\cite{Malik:2003mv}
\bibitem{Malik:2003mv}
  K.~A.~Malik and D.~Wands,
  %``Evolution of second order cosmological perturbations,''
  Class.\ Quant.\ Grav.\  {\bf 21}, L65 (2004)
  [arXiv:astro-ph/0307055].
  %%CITATION = ASTRO-PH 0307055;%%

%\cite{Noh:2004bc}
\bibitem{Noh:2004bc}
  H.~Noh and J.~c.~Hwang,
  %``Second-order perturbations of the Friedmann world model,''
  Phys.\ Rev.\ D {\bf 69}, 104011 (2004).
  %%CITATION = PHRVA,D69,104011;%%

\bibitem{Vernizzi:2004nc}
  F.~Vernizzi,
  %``On the conservation of second-order cosmological perturbations in a  scalar
  %field dominated Universe,''
  Phys.\ Rev.\ D {\bf 71}, 061301 (2005)
  [arXiv:astro-ph/0411463].
  %%CITATION = ASTRO-PH 0411463;%%

%\cite{Tomita:2005et}
\bibitem{Tomita:2005et}
  K.~Tomita,
   %``Relativistic second-order perturbations of nonzero-Lambda flat
  %cosmological models and CMB anisotropies,''
  Phys.\ Rev.\ D {\bf 71}, 083504 (2005)
  [arXiv:astro-ph/0501663].
  %%CITATION = ASTRO-PH 0501663;%%

%%%%%%%%%%%%%%%%%%%%%%%%%%%%%%%%%%%%%%%%%%%%%%%%%%%%%%%%%%%%%%%%%%%%%%%%%%%%%
%nonlinear
% \cite{Salopek:1990jq,Comer:1994np,Deruelle:1994iz,Sasaki:1998ug,Lyth:2003im,Rigopoulos:2003ak,Rigopoulos:2004gr,Rigopoulos:2005xx,Rigopoulos:2005,Lyth:2004gb}

\bibitem{Salopek:1990jq}
  D.~S.~Salopek and J.~R.~Bond,
  %``Nonlinear Evolution Of Long Wavelength Metric Fluctuations In Inflationary
  %Models,''
  Phys.\ Rev.\ D {\bf 42}, 3936 (1990).

\bibitem{Comer:1994np}
  G.~L.~Comer, N.~Deruelle, D.~Langlois and J.~Parry,
  %``Growth or decay of cosmological inhomogeneities as a function of their
  %equation of state,''
  Phys.\ Rev.\ D {\bf 49} (1994) 2759.
  %%CITATION = PHRVA,D49,2759;%%

\bibitem{Deruelle:1994iz}
  N.~Deruelle and D.~Langlois,
  %``Long wavelength iteration of Einstein's equations near a space-time
  %singularity,''
  Phys.\ Rev.\ D {\bf 52}, 2007 (1995)
  [arXiv:gr-qc/9411040].
  %%CITATION = GR-QC 9411040;%%

%\cite{Sasaki:1998ug}
\bibitem{Sasaki:1998ug}
  M.~Sasaki and T.~Tanaka,
  %``Super-horizon scale dynamics of multi-scalar inflation,''
  Prog.\ Theor.\ Phys.\  {\bf 99}, 763 (1998)
  [arXiv:gr-qc/9801017].
  %%CITATION = GR-QC 9801017;%%

%\cite{Lyth:2003im}
\bibitem{Lyth:2003im}
  D.~H.~Lyth and D.~Wands,
  %``Conserved cosmological perturbations,''
  Phys.\ Rev.\ D {\bf 68}, 103515 (2003)
  [arXiv:astro-ph/0306498].
  %%CITATION = ASTRO-PH 0306498;%%

%\cite{Rigopoulos:2003ak}
\bibitem{Rigopoulos:2003ak}
  G.~I.~Rigopoulos and E.~P.~S.~Shellard,
   %``The separate universe approach and the evolution of nonlinear  superhorizon
  %cosmological perturbations,''
  Phys.\ Rev.\ D {\bf 68}, 123518 (2003)
  [arXiv:astro-ph/0306620].
  %%CITATION = ASTRO-PH 0306620;%%

%\cite{Rigopoulos:2004gr}
\bibitem{Rigopoulos:2004gr}
  G.~I.~Rigopoulos and E.~P.~S.~Shellard,
  %``Non-linear inflationary perturbations,''
  JCAP {\bf 0510}, 006 (2005)
  [arXiv:astro-ph/0405185].
  %%CITATION = ASTRO-PH 0405185;%%


%\cite{Rigopoulos:2005xx}
\bibitem{Rigopoulos:2005xx}
  G.~I.~Rigopoulos, E.~P.~S.~Shellard and B.~W.~van Tent,
  %``Non-linear perturbations in multiple-field inflation,''
  Phys.\ Rev.\ D {\bf 73}, 083521 (2006)
  [arXiv:astro-ph/0504508].
  %%CITATION = ASTRO-PH 0504508;%%

\bibitem{Rigopoulos:2005}
  G.~I.~Rigopoulos, E.~P.~S.~Shellard and B.~W.~van Tent,
  %``Large non-Gaussianity in multiple-field inflation,''
  Phys.\ Rev.\ D {\bf 73}, 083522 (2006)
  [arXiv:astro-ph/0506704].
  %%CITATION = ASTRO-PH 0506704;%%

%\cite{Lyth:2004gb}
\bibitem{Lyth:2004gb}
  D.~H.~Lyth, K.~A.~Malik and M.~Sasaki,
  %``A general proof of the conservation of the curvature perturbation,''
  JCAP {\bf 0505}, 004 (2005)
  [arXiv:astro-ph/0411220].
  %%CITATION = ASTRO-PH 0411220;%%

%%%%%%%%%%%%%%%%%%%%%%%%%%%%%%%%%%%%%%%%%%%%%%%%%%%%%%%%%%%%%%%%%%%%%%%%%%%%%
% separate universe
%\cite{Lyth:2005fi}
\bibitem{Lyth:2005fi}
D.~H.~Lyth and Y.~Rodriguez,
  %``Non-gaussianity from the second-order cosmological perturbation,''
  Phys.\ Rev.\  D {\bf 71}, 123508 (2005)
  [arXiv:astro-ph/0502578];
  %%CITATION = PHRVA,D71,123508;%%
{\em ibid.},
  %``The inflationary prediction for primordial non-gaussianity,''
  Phys.\ Rev.\ Lett.\  {\bf 95}, 121302 (2005)
  [arXiv:astro-ph/0504045].
  %%CITATION = ASTRO-PH 0504045;%%
%%%%%%%%%%%%%%%%%%%%%%%%%%%%%%%%%%%%%%%%%%%%%%%%%%%%%%%%%%%%%%%%%%%%%%%%%%%%%%


%\cite{Ellis:1989jt}
\bibitem{Ellis:1989jt}
  G.~F.~R.~Ellis and M.~Bruni,
  %``Covariant And Gauge Invariant Approach To Cosmological Density
  %Fluctuations,''
  Phys.\ Rev.\ D {\bf 40}, 1804 (1989).
  %%CITATION = PHRVA,D40,1804;%%

%\cite{Hawking:1966qi}
\bibitem{Hawking:1966qi}
  S.~W.~Hawking,
  %``Perturbations of an expanding universe,''
  Astrophys.\ J.\  {\bf 145}, 544 (1966).
  %%CITATION = ASJOA,145,544;%%



\bibitem{Bruni:1991kb}
  M.~Bruni, G.~F.~R.~Ellis and P.~K.~S.~Dunsby,
  %``Gauge invariant perturbations in a scalar field dominated universe,''
  Class.\ Quant.\ Grav.\  {\bf 9}, 921 (1992).
  %%CITATION = CQGRD,9,921;%%



\bibitem{Gordon:2000hv}
  C.~Gordon, D.~Wands, B.~A.~Bassett and R.~Maartens,
  %``Adiabatic and entropy perturbations from inflation,''
  Phys.\ Rev.\ D {\bf 63}, 023506 (2001)
  [arXiv:astro-ph/0009131].
  %%CITATION = ASTRO-PH 0009131;%%


%\cite{GrootNibbelink:2000vx}
\bibitem{bartjan}
  S.~Groot Nibbelink and B.~J.~W.~van Tent,
  %``Density perturbations arising from multiple field slow-roll inflation,''
  arXiv:hep-ph/0011325;
  %%CITATION = HEP-PH 0011325;%%
  S.~Groot Nibbelink and B.~J.~W.~van Tent,
  %``Scalar perturbations during multiple field slow-roll inflation,''
  Class.\ Quant.\ Grav.\  {\bf 19}, 613 (2002)
  [arXiv:hep-ph/0107272].
  %%CITATION = HEP-PH 0107272;%%


%\cite{Langlois:1999dw}
\bibitem{Langlois:1999dw}
  D.~Langlois,
  %``Correlated adiabatic and isocurvature perturbations from double
  %inflation,''
  Phys.\ Rev.\ D {\bf 59}, 123512 (1999)
  [arXiv:astro-ph/9906080].
  %%CITATION = ASTRO-PH 9906080;

\bibitem{Malik:2005cy}
  K.~A.~Malik,
  %``Gauge-invariant perturbations at second order: Multiple scalar fields  on
  %large scales,''
  JCAP {\bf 0511}, 005 (2005)
  [arXiv:astro-ph/0506532].
  %%CITATION = ASTRO-PH 0506532;%%

% End Intro
%%%%%%%%%%%%%%%%%%%%%%%%%%%%%%%%%%%%%%%%%%%%%%%%%%%%%%%%%%%%%%%%%%%%%%%%%%%%%%%%%

\bibitem{wald}
  R.~M.~Wald,
  ``General Relativity,'', Chicago University Press, Usa (1984).
%\href{http://www.slac.stanford.edu/spires/find/hep/www?irn=1334239}{SPIRES entry}



%\cite{Sasaki:1995aw}
\bibitem{Sasaki:1995aw}
  M.~Sasaki and E.~D.~Stewart,
  %``A General analytic formula for the spectral index of the density
  %perturbations produced during inflation,''
  Prog.\ Theor.\ Phys.\  {\bf 95}, 71 (1996)
  [arXiv:astro-ph/9507001].
  %%CITATION = ASTRO-PH 9507001;%%

\bibitem{Starobinsky}
  A.~A.~Starobinsky,
%   ``MULTICOMPONENT DE SITTER (INFLATIONARY) STAGES AND THE GENERATION OF
  %PERTURBATIONS,''
  JETP Lett.\  {\bf 42}, 152 (1985)
  [Pisma Zh.\ Eksp.\ Teor.\ Fiz.\  {\bf 42}, 124 (1985)].
  %%CITATION = JTPLA,42,152;%%

\bibitem{SM}
M.~Sasaki, Prog.~Theor.~Phys.~{\bf 76}, 1036 (1986);
V.~F.~Mukhanov, Zh.~\'Eksp.~Teor.~Fiz.~{\bf 94}, 1 (1988)
[Sov.~Phys.~JETP {\bf 68}, 1297 (1988)].

\bibitem{TN}
A.~Taruya and Y.~Nambu, Phys.~Lett.~{\bf B428}, 37 (1998).


%\cite{Bardeen:1980kt}
\bibitem{Bardeen:1980kt}
   J.~M.~Bardeen,
   %``Gauge Invariant Cosmological Perturbations,''
   Phys.\ Rev.\ D {\bf 22}, 1882 (1980).
   %%CITATION = PHRVA,D22,1882;%%


%\cite{Bardeen:1983qw}
\bibitem{Bardeen:1983qw}
  J.~M.~Bardeen, P.~J.~Steinhardt and M.~S.~Turner,
  %``Spontaneous Creation Of Almost Scale - Free Density Perturbations In An
  %Inflationary Universe,''
  Phys.\ Rev.\ D {\bf 28}, 679 (1983).


%---------------------------------------------------------------------------


\bibitem{Enqvist:2006fs}
K.~Enqvist, J.~Hogdahl, S.~Nurmi and F.~Vernizzi,
%``A covariant generalization of cosmological perturbation theory,'' 
Phys.\ Rev.\ D {\bf 75}, 023515 (2007) 
[arXiv:gr-qc/0611020].



% change FV 09-01-2007
\bibitem{Finelli:2006wk}
  F.~Finelli, G.~Marozzi, G.~P.~Vacca and G.~Venturi,
  %``Second order gauge-invariant perturbations during inflation,''
  Phys.\ Rev.\ D {\bf 74}, 083522 (2006)
  [arXiv:gr-qc/0604081].
  %%CITATION = GR-QC 0604081;%%


%\cite{Nakamura:2006rk}
\bibitem{Nakamura}
  K.~Nakamura,
  %``Gauge-invariant formulation of the second-order cosmological
  %perturbations,''
  Phys.\ Rev.\ D {\bf 74}, 101301 (2006)
  [arXiv:gr-qc/0605107];
    ibid.,
  %``Second-order gauge invariant cosmological perturbation theory: Einstein
  %equations in terms of gauge invariant variables,''
  arXiv:gr-qc/0605108.
  %%CITATION = GR-QC 0605108;%%
% end of change





%\cite{Enqvist:2004bk}
\bibitem{Enqvist:2004bk}
  K.~Enqvist and A.~Vaihkonen,
  %``Non-Gaussian perturbations in hybrid inflation,''
  JCAP {\bf 0409}, 006 (2004)
  [arXiv:hep-ph/0405103].
  %%CITATION = HEP-PH 0405103;%%

\bibitem{Anupam}
  A.~Jokinen and A.~Mazumdar,
  % ``Very Large Primordial Non-Gaussianity from multi-field: Application to
  %Massless Preheating,''
  JCAP {\bf 0604}, 003 (2006)
  [arXiv:astro-ph/0512368].
  %%CITATION = ASTRO-PH 0512368;%%


\bibitem{Cline}
  N.~Barnaby and J.~M.~Cline,
  % ``Nongaussian and nonscale-invariant perturbations from tachyonic preheating
  %in hybrid inflation,''
  Phys.\ Rev.\ D {\bf 73}, 106012 (2006)
  [arXiv:astro-ph/0601481].
  %%CITATION = ASTRO-PH 0601481;%%



%\cite{Vernizzi:2006ve}
\bibitem{Vernizzi:2006ve}
  F.~Vernizzi and D.~Wands,
  %``Non-Gaussianities in two-field inflation,''
  JCAP {\bf 0605}, 019 (2006)
  [arXiv:astro-ph/0603799].
  %%CITATION = ASTRO-PH 0603799;%%


\bibitem{Seery:2006vu}
D.~Seery, J.~E.~Lidsey and M.~S.~Sloth,
    %``The inflationary trispectrum,''
    arXiv:astro-ph/0610210.
    %%CITATION = ASTRO-PH 0610210;%%



%%%%%%%%%%%%%%%%%%%%%%%%%%%%%%%%%%%%%%%%%%%%%%%%%%%%%%%%%%%%%%%%%%%%%%%%%%%%%









\end{thebibliography}
\end{document}